\documentclass[%
 reprint,
superscriptaddress,
 amsmath,amssymb,
 aps,
]{revtex4-1}

\usepackage{graphicx}
\usepackage{dcolumn}
\usepackage{bm}
\usepackage[mathlines]{lineno}
\usepackage{hyperref}
\usepackage{xcolor}
\hypersetup{
    colorlinks=true,
    linkcolor=blue,
    citecolor=blue,
    urlcolor=blue,
}
\usepackage{color,soul}

\usepackage[T1]{fontenc}
\usepackage{siunitx}
\DeclareSIUnit\torr{Torr}
\DeclareSIUnit\oersted{Oe}

\begin{document}

\preprint{APS/123-QED}

\title{Energy-efficient W$_{100-x}$Ta$_x$/CoFeB/MgO spin Hall nano-oscillators}

\author{Nilamani Behera}
\affiliation{Physics Department, University of Gothenburg, 412 96 Gothenburg, Sweden.}
\author{Himanshu Fulara}
\email{himanshu.fulara@ph.iitr.ac.in}
\affiliation{Department of Physics, Indian Institute of Technology Roorkee, Roorkee 247667, India}
\author{Lakhan Bainsla}
\affiliation{Physics Department, University of Gothenburg, 412 96 Gothenburg, Sweden.}
\author{Akash Kumar}
\affiliation{Physics Department, University of Gothenburg, 412 96 Gothenburg, Sweden.}
\author{Mohammad Zahedinejad}
\affiliation{Physics Department, University of Gothenburg, 412 96 Gothenburg, Sweden.}
\author{Afshin Houshang}
\affiliation{Physics Department, University of Gothenburg, 412 96 Gothenburg, Sweden.}
\author{Johan \AA kerman}
\email{johan.akerman@physics.gu.se}
\affiliation{Physics Department, University of Gothenburg, 412 96 Gothenburg, Sweden.}




\date{\today}

\begin{abstract}
We investigate a W--Ta alloying route to 
reduce the auto-oscillation current densities and the power consumption of nano-constriction based spin Hall nano-oscillators. 
Using spin-torque ferromagnetic resonance (ST-FMR) measurements 
on microbars of W$_{\text{100-x}}$Ta$_{\text{x}}$(5 nm)/CoFeB(t)/MgO stacks with $t=$ 1.4, 1.8, and 2.0 nm, we measure a substantial improvement in both the spin-orbit torque efficiency and the spin Hall conductivity. 
We demonstrate a 34\% reduction in threshold auto-oscillation current density, 
which translates into a 
64\% reduction in 
power consumption as compared to pure W based SHNOs. Our work demonstrates the promising aspects of W--Ta alloying 
for the energy-efficient operation of emerging spintronic devices.

\end{abstract}

\pacs{Valid PACS appear here}
\maketitle                         
\section{\label{sec:level1}Introduction}

Current induced spin orbit torques (SOTs) \cite{gambardella2011philtrans,Manchon2019RevModPhys, Shao2021ieeetmag}, originating from the spin Hall effect (SHE) \cite{Dyakonov1971pl,Hirsch1999prl,Sinova2015RevModPhys} in a non magnetic heavy metal (HM)/ferromagnet(FM) heterostructure, have recently emerged as a promising energy-efficient route for next-generation ultra-fast spintronic devices such as spin Hall nano-oscillators (SHNOs) \cite{Demidov2012b, liu2012prl, Demidov2014apl, Ranjbar2014, Chen2016procieee,  Zahedinejad2018apl,Haidar2019natcomm,durrenfeld2017nanoscale, Duan2014b}, non-volatile SOT-based magnetic random access memory (SOT-MRAM) \cite{liu2012science, Miron2011nat}, SOT driven magnonics \cite{Xing2017prappl,Demidov2020jap}, and spin logic devices \cite{Dieny2020natelec}. The magnetization dynamics in such devices is driven by a pure spin current, which exerts an anti-damping torque on the local magnetization vector of the adjacent FM layer, resulting in magnetization switching in SOT-MRAM or auto-oscillations in SHNOs.

Nano-constriction SHNOs \cite{Demidov2014apl,Awad2020apl} currently receive an increasing interest as they demonstrate rich non-linear magnetodynamics \cite{Zahedinejad2017ieeeml,Dvornik2018prappl,Spicer2018prb,mazraati2018prappl,Hache2019apl,Smith2020prb,divinskiy2017advm,Fulara2019SciAdv,fulara2020natcomm}, can be implemented using a wide range of materials \cite{Mazraati2016apl,Evelt2018scirep, Hache2020apl, Haidar2021apl,Sato2019prl,Safranski2019natnano,Chen2020commphys}, and show a great propensity for mutual synchronization \cite{Awad2016natphys,Zahedinejad2020natnano}, which leads to orders of magnitude higher signal coherence \cite{Zahedinejad2020natnano} and allow for oscillator based neuromorphic computing \cite{Zahedinejad2020natnano,Singh2021aipadv, zahedinejad2020arxiv,Garg2021neurcomp} and Ising Machines \cite{Albertsson2021apl,houshang2020arxiv}. 

However, a key challenge remains to minimize the switching and auto-oscillation threshold current densities and the associated energy consumption. 
There has been tremendous effort to enhance the spin Hall angle ($\mathit{\theta_{SH}}$) and the SOT efficiency ($\mathit{\xi_{SOT}}$) through different routes such as employing different material combinations \cite{Pai2012}, incorporating oxygen \cite{Demasius2016natcomm,utkarsh2021apl}, and dusting the HM surface with Hf \cite{nguyen2015apl,mazraati2018apl}. Even so, an increase in $\mathit{\xi_{SOT}}$ is typically achieved at the expense of an equivalent decrease in the longitudinal conductivity ($\sigma_{0}$) leading to high power dissipation during device operation. A better figure of merit is therefore the spin Hall conductivity: $\mathit{\sigma_{SH}} = \sigma_{0} \mathit{\xi_{SOT}} \hbar/2e$. 

Recent studies have shown that alloying of HMs is a promising route to tune both  $\mathit{\xi_{SOT}}$ and $\mathit{\sigma_{SH}}$ \cite{obstbaum2016prl,sui2017prb,zhu2018pra,kim2020apl}. In 2018, Zhu et al.~\cite{zhu2018pra} reported highly efficient spin current generation in a Au$_{0.25}$Pt$_{0.75}$ alloy exhibiting a relatively low longitudnial resistivity ($\rho_0$) of approximately 83 $\mu\Omega$.cm and large $\mathit{\xi_{SOT}}$ = 0.35 in bilayers with Co. Quite recently, Kim et al.~showed an enhancement of $\mathit{\sigma_{SH}}$ in W-Ta alloys using spin transfer ferromagnetic resonance (ST-FMR) measurements \cite{kim2020apl}. While a reasonably high $\mathit{\theta_{SH}}$ = –0.3 with a relatively lower $\rho_0$ of 100 $\mu\Omega$.cm was reported at a 11{$\%$} Ta concentration, its ultimate effect on $\mathit{\sigma_{SH}}$ and its potential for reducing the threshold current and the associated energy consumption of SHNOs have yet to be investigated. 

Here, we report on an extensive study of W--Ta alloying in 
W$_{100-x}$Ta$_{x}$(5 nm)/Co$_{20}$Fe$_{60}$B$_{20}$(t)/MgO(2 nm) nano-constriction SHNOs with $t=$ 1.4, 1.8, and 2.0 nm. 
W--Ta alloying results in a simultaneous improvement of both $\mathit{\xi_{SOT}}$ and $\mathit{\sigma_{SH}}$, 
which translates into substantial reductions in the threshold current density ($I_{th}$) and the power consumption. 
Using ST-FMR measurements on microbars 
we first study the magnetodynamical properties to extract $\xi_{SOT}$ and $\mathit{\sigma_{SH}}$. $\xi_{SOT}$ first strengthens to an optimum value of -0.61 at 10\% Ta, 
compared to -0.46 for pure $\beta${-W}, and then weakens to an efficiency of -0.21 as the Ta content is increased to 25\%. 
As the resistivity drops substantially with Ta content, alloying 
gives rise to a giant $\sim$109{$\%$} increase in $\mathit{\sigma_{SH}}$ at 18\% Ta compared to 
pure $\beta${-W}. Both $\xi_{SOT}$ and $\mathit{\sigma_{SH}}$ scale inversely with the CoFeB thickness, as expected. Finally, we quantify the alloying effect on the auto-oscillation current densities by fabricating SHNOs of two different constriction widths, 50 and 120 nm. The lowest threshold currents are observed for 12\% Ta, with a $\sim$34{$\%$} reduction in auto-oscillation current densities. The reduced threshold current densities translate into a 64{$\%$ reduced power consumption as compared to pure W based SHNOs. The trade-off between SOT efficiency, resistivity, and equivalent SHC demonstrates the promising aspects of W--Ta alloying approach for energy-efficient and CMOS compatible operation of emerging spintronic devices.

\begin{figure*}
\includegraphics[width=16cm]{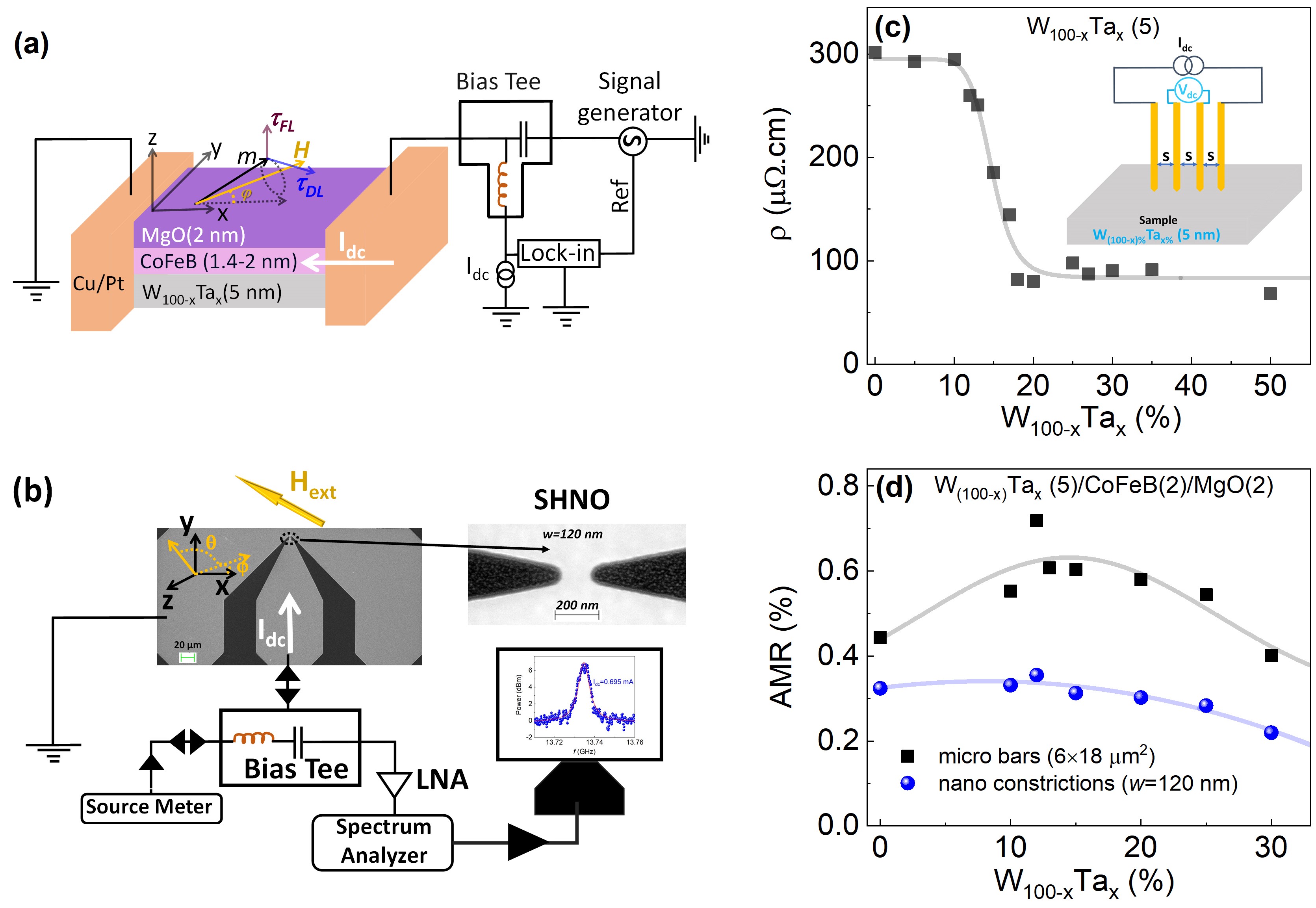}
\caption{ Device schematic, resistivity, and magnetoresistance. (a) Schematic of ST-FMR measurement set-up (b) Schematic of auto-oscillation measurement set-up along with Scanning electron micrograph (SEM) image of a 120 nm nano-constroction width SHNO (c) Variation of resistivity of W-Ta alloy thin films as a function of Ta composition. Solid line shows the average behaviour of resistivity variation with W-Ta alloy composition (d) Anisotropic magnetoresitance (AMR) measured as a function of different alloy W-Ta alloy composition on microbars (black squares) and 120 nm nano-constriction SHNO (blue dots). Solid line shows the average behaviour of magnetoresistance variation with W-Ta alloy composition. 
} 
\label{fig:1}
\end{figure*}

\subsection{\label{sec:level2}Experimental Details}
W$_{100-x}$Ta$_{x}$(5 nm)/Co$_{20}$Fe$_{60}$B$_{20}$($t$)/MgO(2 nm) material stacks, with $t=$ 1.4, 1.8, and 2.0 nm, 
were deposited on highly resistive Si(100) substrates using an AJA Orion-8 magnetron sputtering system, working at a base pressure of 3 $\times$10$^{-8}$ Torr, while the Argon pressure during sputtering was maintained at 3 mTorr for all layers. A 4 nm SiO$_{2}$ capping layer was added to protect the MgO from moisture. The W$_{100-x}$Ta$_{x}$ alloy films were grown by  co-sputtering W and Ta metal targets subjected to dc and rf power, respectively. 
During deposition, the growth rate of W was kept low at 0.1 Å/s to obtain the desired $\beta$-phase with a high spin Hall angle \cite{Zahedinejad2018apl,Pai2012}. The growth rates of Co$_{20}$Fe$_{60}$B$_{20}$, MgO, and SiO$_{2}$ layer were maintained at 0.13 Å/s, 0.06 Å/s, and 0.08 Å/s, respectively, for good thickness control. The stacks were subsequently annealed at 300~$^{\circ}$C for 60 min at the chamber's base pressure to crystallize MgO at the interface as well as CoFeB. Separate individual stacks of of pure W, W$_{100-x}$Ta$_{x}$ alloys, and the CoFeB layers were carried out for their respective resistivity measurements. 

To fabricate nano-constriction SHNOs, the sample stack surface was covered with negative electron resist (HSQ) followed by an exposure to electron beam lithography (RAITH EBPG 5200 EBL). Nano-constrictions of 50 and 120 nm were defined in 4$\times$12$~\mu m^2$  mesas. Furthermore, 6$\times$18$~\mu m^2$  and 6$\times$12$~\mu m^2$  micro bars were also designed to characterize the stacks using spin-torque-induced ferromagnetic resonance (ST-FMR) measurements. 
Subsequently, these defined patterns were transferred to the stack by Ar ion beam etching using an Oxford Ionfab 300 Plus etcher. Later, the negative resist was removed, and optical lift-off lithography was carried out to define ground–signal–ground (GSG) coplanar waveguides (CPW) of a thick Cu(800 nm)/Pt(20 nm) bilayer. To ensure a good electrical contact between the CPW and the SHNOs, the MgO/SiO$_{x}$ layers were removed in the CPW defined area by substrate plasma cleaning at an rf power of 40 W in the AJA Orion-8 sputtering chamber right before Cu/Pt deposition.

The spin transport measurements on the micro bar devices were carried out using a room-temperature ST-FMR setup with a fixed in-plane angle, $\phi$=30$^{\circ}$ to estimate the SOT efficiencies and other magneto-dynamical parameters. Figure ~\ref{fig:1}a schematically illustrates the ST-FMR measurement details on the patterned 6 $\mu$m width bars of all W$_{100-x}$Ta$_{x}$(5 nm)/CoFeB(1.4-2 nm)/MgO(2) stacks. The microwave current, $\mathit{I_{rf}}$, was modulated at 98.76 Hz, and injected into the microbars through a high frequency bias Tee, producing spin-orbit torques and Oersted field under the presence of out-of-plane (OOP) magnetic field. The Oersted field generates an OOP torque on the CoFeB magnetization ($\mathit \tau_{Oe}$) and the additional field-like torque ($\mathit \tau_{FLT}$ ) and damping-like torque ($\mathit \xi_{DLT}$}) are induced due to exchange interaction of the transverse spin current density with the magnetization  in the CoFeB layer \cite{skinner2014apl}. These torques govern the magnetization dynamics in the CoFeB layer, which results in an oscillatory change in the resistance of the device due to anisotropic magetoresistance (AMR) and spin Hall magnetoresistance (SMR) of the CoFeB layer. The oscillating resistance mixes with $\mathit{I_{rf}}$ and produces a dc voltage ($\mathit{V_{mix}}$) across the microbar, which is detected on the modulating frequency using a lock-in amplifier. 

All magnetization auto-oscillation measurements were performed using a custom-built probe station with the sample mounted at a fixed in-plane angle on an OOP rotatable sample holder lying between the pole pieces of an electromagnet generating a uniform magnetic field. Figure ~\ref{fig:1}b schematically shows the experimental set-up for auto-oscillation measurements on a SHNO device of width 120 nm. Here, a positive direct current was fed to the SHNO device through the dc port of the high frequency bias Tee under a fixed OOP magnetic field and the resulting auto-oscillation signal was first amplified by a low noise amplifier of gain 72 dB and thereafter recorded using a Rohde \& Schwarz (10 Hz to 40 GHz) spectrum analyzer with a low-resolution bandwidth of 300 kHz.

\subsection{Results and Discussion}
Sheet resistance measurements were carried out on W$_{100-x}$Ta$_{x}$(5 nm) alloy thin films to determine their resistivity \emph{vs.}~Ta concentration. Fig.~\ref{fig:1}(c), showing a plot of $\rho$  \emph{vs.}~
\%Ta, 
reveals that the resistivity starts out at 
300~$\mu\Omega$.cm, which is the typical resistivity for A15 type of materials \cite{Pai2012,petroff1973microstructure}, and then exhibits a steep drop, between 10\% and 18\% Ta, to about 90~$\mu\Omega$.cm for all higher concentrations, \emph{i.e.}~a greater than 3x difference. Figure 1d shows the W$_{100-x}$Ta$_{x}$ composition dependent AMR behavior of 6$\times$18$~\mu m^2$ microbars and 120 nm nanoconstrctions. The AMR shows a maximum at 12\% Ta 
as compared to pure W and then trends downward with further increase of Ta concentration.

\begin{figure}
\includegraphics[width=8.cm]{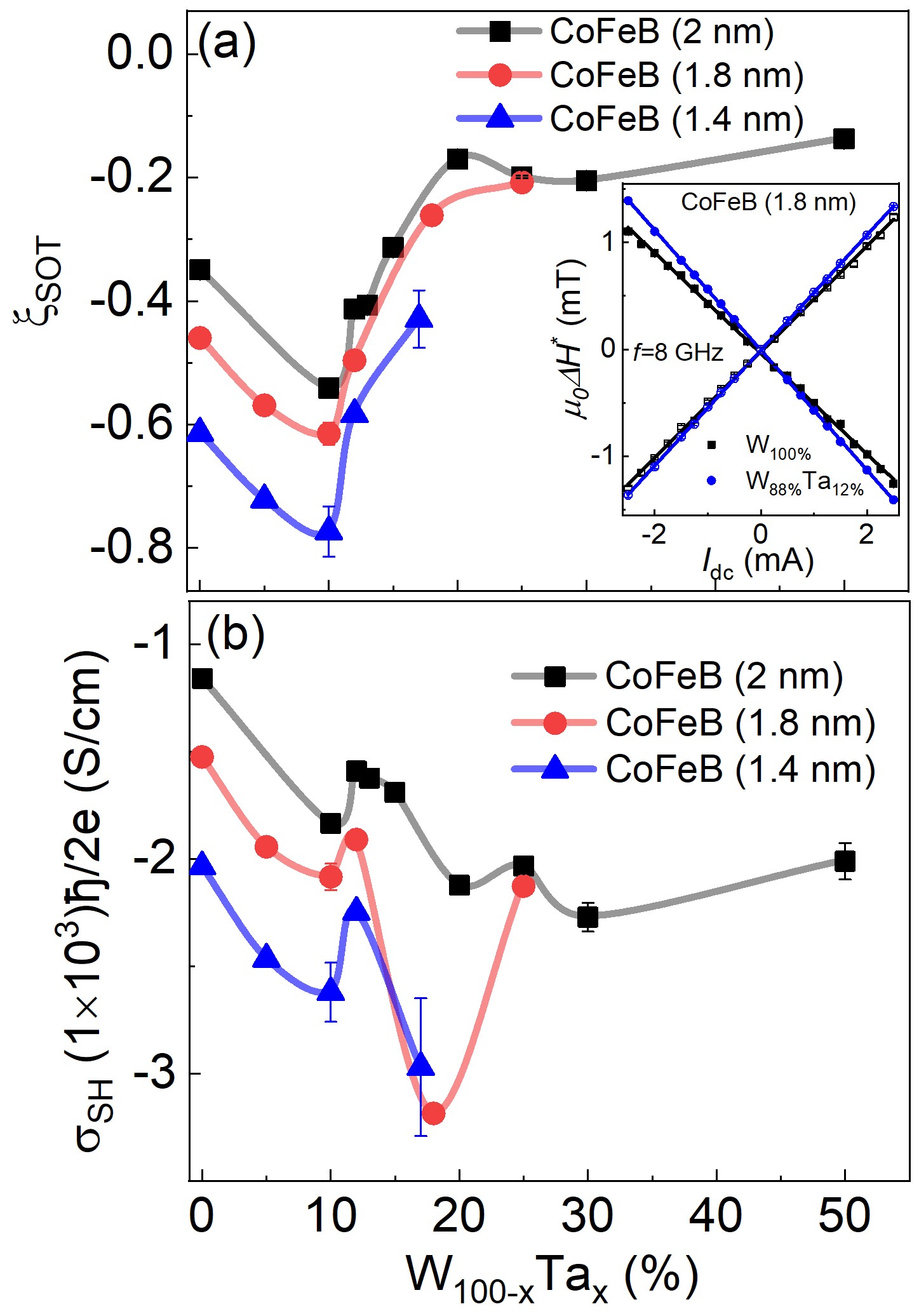}
\caption{ST-FMR measurements performed on W$_{100-x}$Ta$_{x}$(5)/Co$_{20}$Fe$_{60}$B$_{20}$($t_{CoFeB}$ = 1.4, 1.8 and 2 nm)/MgO(2) microbars.
(a) Dependence of SOT efficiency ($\xi_{SOT}$) on W-Ta alloy composition for three different CoFeB thicknesses exhibiting similar behaviour. Solid line shows the average behaviour. Inset shows linewidth dependence on dc current to estimate SOT efficiency. (b) Variation of spin Hall conductivity ($\sigma_{SH}$) as a function of W-Ta alloy composition shown for three different CoFeB thicknesses. 
}
\label{fig:2}
\end{figure}

\begin{figure*}
\includegraphics[width=12cm]{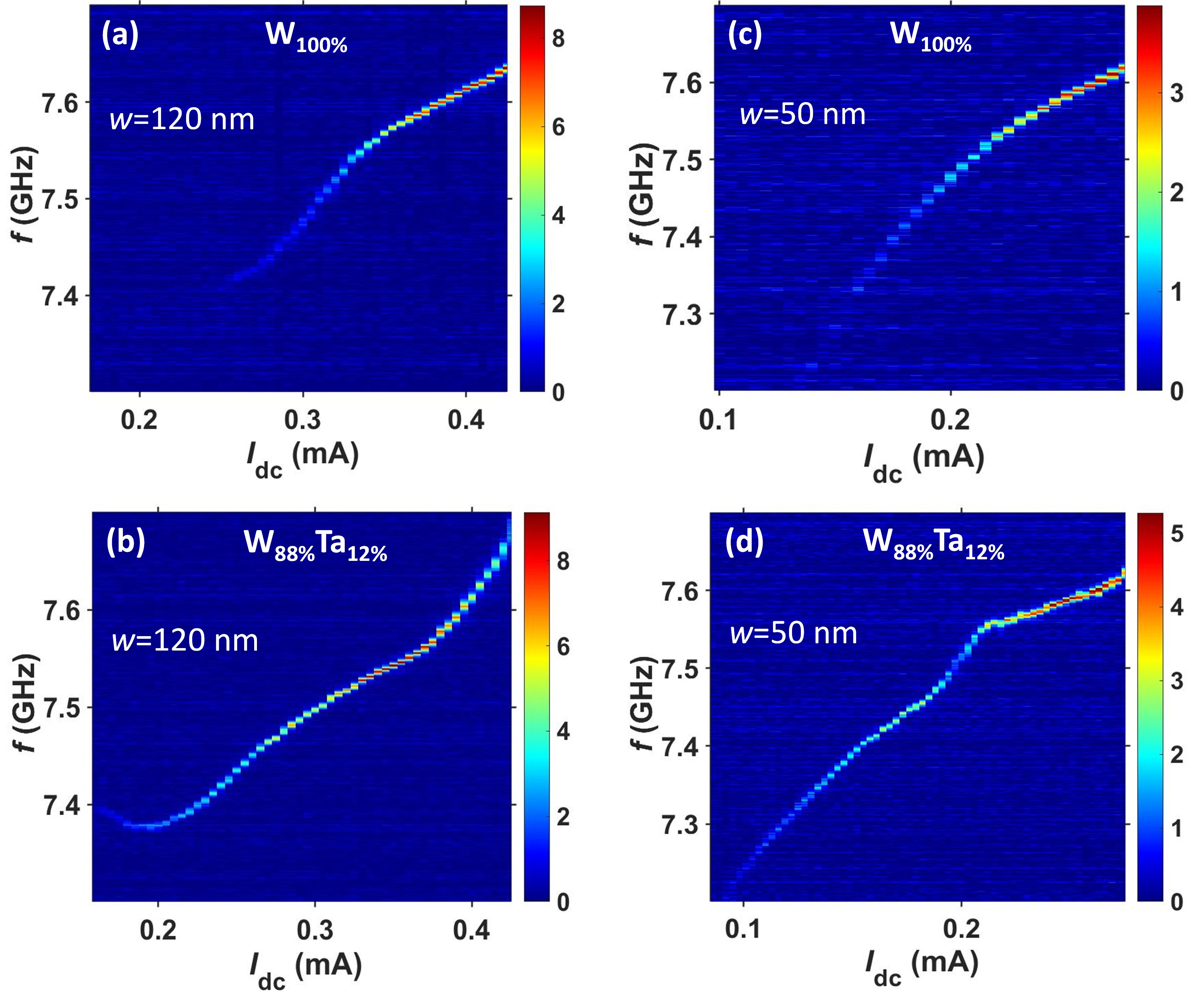}
\caption{PSDs of the auto-oscillations vs. current showing a consistent drop in threshold currents of W-Ta alloy based SHNOs comprising CoFeB thickness $t_{CoFeB}$ = 1.4 nm, measured at out-of-plane magnetic field, $\mu_{0}$H=0.3 T, $\theta$=60$^{\circ}$, $\phi$=22$^{\circ}$.}
\label{fig:3}
\end{figure*}

STFMR measurements 
of all W$_{100-x}$Ta$_{x}$(5)/CoFeB(1.4-2)/MgO(2) stacks were carried out to estimate the spin orbit torque efficiency, $\xi_{SOT}$=$\frac{2e}{\hbar}(J_{s}/J_{c})$, i.e.~charge to spin current conversion efficiency. The observed STFMR spectra were fitted with 
a sum of symmetric ($F_{S}$($\it{H}$)) and anti-symmetric ($F_{A}$($\it{H}$)) Lorentizans 
\cite{Liu2011prl,Demasius2016natcomm} as $V_{mix}$=$V_{0}$[S$F_{S}$($\it{H}$)+A$F_{A}$($\it{H}$)], where $V_{0}$ is the amplitude of the mixing voltage and S and A are the symmetric and anti-symmetric Lorentzian weight factors, accounting for anti-damping and field-like torques, respectively. 
The resonance field ($H_{r}$) and linewidth ($\Delta{H}$) were extracted. and the effective magnetization, $\mu_{0}\mathit{M_{eff}}$, 
was determined from fits of 
$f$ \emph{vs.}~$H_{r}$ 
to the Kittel equation, \(\emph{f}=(\gamma/2\pi)\mu_0\sqrt{(\emph{$H_{r}$+$H_{k}$})(\emph{$H_{r}$}+\emph{$H_{k}$}+M_{eff})}\), by assuming a constant gyromagnetic ratio values for all concentrations of W$_{(100-x)_{\%}}$Ta$_{x_{\%}}$ but different for different 
CoFeB thickness: $\gamma$/2$\pi$ = 29.4, 30.4, and 30.5 GHz/T for 
1.4 nm, 1.8 nm, and 2 nm of CoFeB. 
The Gilbert damping ($\alpha$) was extracted from $\Delta{H}$ vs.~$f$ using \(\Delta\emph{H}=\Delta\emph{H}_0+(2\pi\alpha\emph{f})/\gamma\). The Ta 
concentration dependent 
$\alpha$ and 
$\mu_{0}\mathit{M_{eff}}$ 
are plotted in the supplementary information (SI) file (see SI file, Fig.S1).

The SOT efficiency, $\xi_{SOT}$ is then determined from the dc current dependent STFMR linewidth ($\Delta\emph{H}$ vs.~$I_{dc}$) analysis \cite{Liu2011prl,Pai2012,Demasius2016natcomm}. The inset of Fig.~\ref{fig:2}a shows a plot of $\Delta{H}^{\star}$ vs.~$I_{dc}$ where $\Delta{H}{^\star}$ = ($\Delta{H}(I_{dc}$)-$\Delta{H}(I_{dc}$=0)) varies linearly with $I_{dc}$ with a slope $\delta{\Delta{H}}/\delta{(I_{dc})}$ indicating the strength of the SOT from which we extract 
\cite{Liu2011prl}:

\begin{equation} \label{eq.1}
\xi_{SOT} = \frac{\delta\Delta H/\delta(I_{dc})}{\frac{2\pi f}{\gamma}\frac{sin\phi}{(H+0.5M_{eff})\mu_{0}M_{s}t_{CoFeB}}\frac{\hbar}{2e}}\frac{R_{W-Ta}+R_{CoFeB}}{R_{CoFeB}}A_{c}
\end{equation}
with $\phi$ the azimuthal angle between $I_{dc}$ and $\mu_{0}H$, $\mu_{0}\mathit{M_{eff}}$ the effective magnetization, and $M_{S}$ = 9.31$\times10^5$ the saturation magnetization of the CoFeB layer \cite{Zahedinejad2018apl}, $\hbar$ and $e$, the reduced Planck’s constant and elementary charge, 
R$_{W-Ta}$ and R$_{CoFeB}$ the resistance of the W$_{100-x}$Ta$_{x}$(5) alloys and the CoFeB layer, respectively, and $A_c$ the cross sectional area of the microbars. For the estimation of $\xi_{SOT}$, the resistivity values as shown in Fig.~\ref{fig:1}b are used for W$_{100-x}$Ta$_{x}$(5) alloys and the resistivity value of CoFeB, 64 $\mu\Omega$.cm, is determined from CoFeB thickness dependent conductance behavior (see SI file, Fig.S2) \cite{kim2013natmat,takeuchi2018apl}. 

As expected, we first observe a monotonic strengthening of $\xi_{SOT}$ from -0.35 to -0.62 with decreasing CoFeB layer thickness, highlighting the interfacial nature of SOT. Then, for all three different CoFeB layer thicknesses, the behaviour of $\xi_{SOT}$ remains qualitatively similar with W-Ta alloy composition, first strengthening linearly with Ta content up until 10\%Ta, then abruptly changing character to rapidly weakening with \%Ta until leveling off above 20\%Ta. 
This behaviour is quite consistent with the resistivity dependence on W-Ta composition, as shown in Fig.~\ref{fig:1}c. As shown in Fig.~\ref{fig:2}a, $\xi_{SOT}$ decreases in the region (x=10${\%}$ to 20${\%}$) where the resistivity shows a steep drop, which is possibly due to the formation of a mixed ($\alpha+\beta$) W phase due to increased Ta alloying. 

We also estimate the spin Hall conductivity 
using the relation  $\xi_{SOT}$=(2e⁄$\hbar$)($\sigma_{SH}$⁄$\sigma_{0}$).
Figure ~\ref{fig:2}b shows the dependence of $\sigma_{SH}$ on W--Ta alloy composition, indicating a qualitatively similar behaviour for all CoFeB thicknesses. We observe the strongest $\sigma_{SH}$ of -3186 ($\hbar$/2e) (S/cm) at 18\% Ta, 
which is a $\sim$109${\%}$ increase compared to 
pure W, 
higher than earlier reported values \cite{kim2020apl,qian2020spin}, but still lower than the theoretically predicted ultimate values \cite{sui2017prb,derunova2019giant}.

\begin{figure*}
\includegraphics[width=16cm]{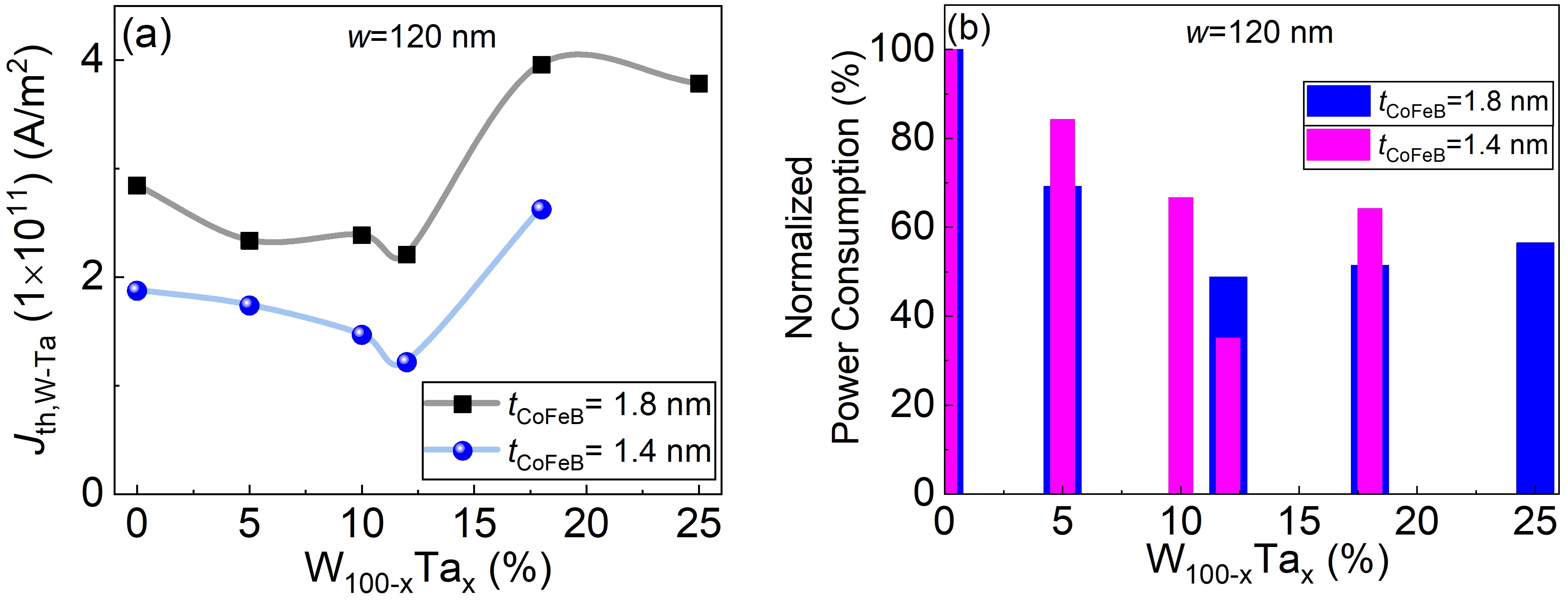}
\caption{(a) 
Variation of threshold current density in W$_{100-x}$Ta$_{x}$ layer as a function of W-Ta alloy composition, extracted for two different CoFeB thicknesses. (b) Dependence of normalized threshold power consumption in a 120 nm SHNO device on W-Ta alloy composition, showing a large 64 \% drop in power.}
\label{fig:4}
\end{figure*}

Next, we quantify how the observed 
improvements in $\xi_{SOT}$ and $\sigma_{SH}$ can eventually lead to energy savings in practical devices. We first performed auto-oscillation measurements on nano-constriction SHNOs fabricated from identical W$_{100-x}$Ta$_{x}$(5 nm)/CoFeB($t$ = 1.4--2 nm)/MgO(2 nm) stacks. Figure ~\ref{fig:3}(a-b) shows the current dependent power spectral density (PSD) plots, measured under a fixed OOP field of 0.3 T, for a 120 nm wide SHNO. The non-monotonic current dependence of the auto-oscillation frequency is typical for these SHNOs due to localized auto-oscillations governed by negative non-linearity in the constriction region \cite{Dvornik2018prappl}. The localization can be fully mitigated by inducing strong perpendicular magnetic anisotropy (PMA) in the ferromagnetic layer, resulting in the excitation of the truly propagating spin-wave auto-oscillations in the constriction region \cite{Fulara2019SciAdv,fulara2020natcomm}. Notably, as shown in Fig.~\ref{fig:3}a, we observe a large reduction of threshold current 
from $\sim$260 $\mu$A for pure W to $\sim$160 $\mu$A for W$_{88}$Ta$_{12}$, 
clearly demonstrating the direct benefit of improved SOT efficiency and spin Hall conductivity. 
This general trend is observed in other nano-constriction widths as shown in Fig.~\ref{fig:3}(c-d), where a 50 nm nano-constriction 
shows a similar large drop in the threshold current, reaching the sub-100 $\mu$A regime. Note that we also observe a qualitatively different current tunability of frequency as we measure auto-oscillations for smaller constriction widths. This is because the effective demagnetization field produced by the constriction edges for narrower constriction width is quite substantial and, therefore, the non-linearity remains positive at all operational currents leading to a blue-shifted frequency behaviour in the entire current range \cite{durrenfeld2017nanoscale, Awad2020apl}. 

In Fig.~\ref{fig:4}a, we summarize the behaviour of threshold current densities, $J_{th,W-Ta}$ 
extracted from auto-oscillation measurements for the 120 nm width SHNOs using the method given in ref.\cite{tiberkevich2007apl}. $J_{th,W-Ta}$ initially decreases with increasing Ta composition (0 to 12\%), and after that increases with higher Ta compositions. The observed trend is qualitatively consistent for two different CoFeB thicknesses, as shown in Fig.~\ref{fig:4}a. We observe a considerable 34${\%}$ reduction in $J_{th,W-Ta}$ as compared to pure W. Further, we estimate the effective power consumption to drive W--Ta based SHNOs using \(\it{P}=\it{(I_{th})}^2R\), where R is the resistance of the SHNO, and normalized by the power consumption in pure W-based SHNOs. Figure~\ref{fig:4}b shows a bar plot comparison of normalized power consumption for the two different CoFeB thicknesses, showing a trend of power consumption similar to Fig.~\ref{fig:4}a. The normalized power consumption shows a minimum value of about 36${\%}$ at 12${\%}$ Ta for 1.4 nm CoFeB thickness, indicating an overall large 64${\%}$ reduction in effective power consumption as a result of W--Ta alloying.

Note that we obtained the most reduction in power consumption at a 12${\%}$ Ta concentration with values of $\sigma_{SH}$ = -2247 ($\hbar$/2e) (S/cm) and $\xi_{SOT}$ = -0.58, despite a minor change in resitivity from 300~$\mu\Omega$.cm to 260~$\mu\Omega$.cm. We argue that the substituted Ta atoms in the cage of W atoms give rise to a minimal distortion of the A15 structure, which can be seen as an apparent increase in $\xi_{SOT}$ value at a 12${\%}$ Ta concentration. However, with increased Ta alloying, W-Ta resistivity shows a steep drop and $\xi_{SOT}$ follows the similar trend. We believe that increased Ta alloying results into the structural changes from $\beta$-W to $\alpha+\beta$-W-Ta phase. Our results evidence that controlled alloying of W with Ta is critical to minimize power consumption in SHNOs and we identified this range as x=10${\%}$ to 20${\%}$.

\subsection{Conclusion}
 We demonstrated a W--Ta alloying route to minimize the auto-oscillation current densities and the power consumption in nano-constriction based SHNOs. The controlled alloying of W with Ta results in a considerable drop in the resistivity, which not only increases the SOT efficiency but also significantly enhances the spin Hall conductivity of the W--Ta layer. As a direct consequence, we observe a large 34${\%}$ drop in threshold current density of SHNOs. The estimated effective power consumption further drops by 64${\%}$, indicating the tremendous potential of the alloying approach over other methods to reduce the energy consumption in emerging spintronic devices. Further down scaling of the constriction width to 20 nm and below should push 
 SHNO operation into the tens of $\mu$A regime. Our experimental results on SHNO devices not only benefits oscillator-based neuromorphic computing in terms of power consumption but also makes the alloy route attractive for other spintronic applications.
 
 \subsection*{Acknowledgements}
 
 This work was partially supported by the Horizon 2020 research and innovation programme No. 835068 "TOPSPIN". This work was also partially supported by the Swedish Research Council (VR Grant No. 2016-05980) and the Knut and Alice Wallenberg Foundation. 
 

\begin{thebibliography}{61}%
\makeatletter
\providecommand \@ifxundefined [1]{%
 \@ifx{#1\undefined}
}%
\providecommand \@ifnum [1]{%
 \ifnum #1\expandafter \@firstoftwo
 \else \expandafter \@secondoftwo
 \fi
}%
\providecommand \@ifx [1]{%
 \ifx #1\expandafter \@firstoftwo
 \else \expandafter \@secondoftwo
 \fi
}%
\providecommand \natexlab [1]{#1}%
\providecommand \enquote  [1]{``#1''}%
\providecommand \bibnamefont  [1]{#1}%
\providecommand \bibfnamefont [1]{#1}%
\providecommand \citenamefont [1]{#1}%
\providecommand \href@noop [0]{\@secondoftwo}%
\providecommand \href [0]{\begingroup \@sanitize@url \@href}%
\providecommand \@href[1]{\@@startlink{#1}\@@href}%
\providecommand \@@href[1]{\endgroup#1\@@endlink}%
\providecommand \@sanitize@url [0]{\catcode `\\12\catcode `\$12\catcode
  `\&12\catcode `\#12\catcode `\^12\catcode `\_12\catcode `\%12\relax}%
\providecommand \@@startlink[1]{}%
\providecommand \@@endlink[0]{}%
\providecommand \url  [0]{\begingroup\@sanitize@url \@url }%
\providecommand \@url [1]{\endgroup\@href {#1}{\urlprefix }}%
\providecommand \urlprefix  [0]{URL }%
\providecommand \Eprint [0]{\href }%
\providecommand \doibase [0]{http://dx.doi.org/}%
\providecommand \selectlanguage [0]{\@gobble}%
\providecommand \bibinfo  [0]{\@secondoftwo}%
\providecommand \bibfield  [0]{\@secondoftwo}%
\providecommand \translation [1]{[#1]}%
\providecommand \BibitemOpen [0]{}%
\providecommand \bibitemStop [0]{}%
\providecommand \bibitemNoStop [0]{.\EOS\space}%
\providecommand \EOS [0]{\spacefactor3000\relax}%
\providecommand \BibitemShut  [1]{\csname bibitem#1\endcsname}%
\let\auto@bib@innerbib\@empty
\bibitem [{\citenamefont {Gambardella}\ and\ \citenamefont
  {Miron}(2011)}]{gambardella2011philtrans}%
  \BibitemOpen
  \bibfield  {author} {\bibinfo {author} {\bibfnamefont {P.}~\bibnamefont
  {Gambardella}}\ and\ \bibinfo {author} {\bibfnamefont {I.~M.}\ \bibnamefont
  {Miron}},\ }\href {\doibase 10.1098/rsta.2010.0336} {\bibfield  {journal}
  {\bibinfo  {journal} {Philosophical Transactions of the Royal Society A:
  Mathematical, Physical and Engineering Sciences}\ }\textbf {\bibinfo {volume}
  {369}},\ \bibinfo {pages} {3175} (\bibinfo {year} {2011})}\BibitemShut
  {NoStop}%
\bibitem [{\citenamefont {Manchon}\ \emph {et~al.}(2019)\citenamefont
  {Manchon}, \citenamefont {\ifmmode~\check{Z}\else \v{Z}\fi{}elezn\'y},
  \citenamefont {Miron}, \citenamefont {Jungwirth}, \citenamefont {Sinova},
  \citenamefont {Thiaville}, \citenamefont {Garello},\ and\ \citenamefont
  {Gambardella}}]{Manchon2019RevModPhys}%
  \BibitemOpen
  \bibfield  {author} {\bibinfo {author} {\bibfnamefont {A.}~\bibnamefont
  {Manchon}}, \bibinfo {author} {\bibfnamefont {J.}~\bibnamefont
  {\ifmmode~\check{Z}\else \v{Z}\fi{}elezn\'y}}, \bibinfo {author}
  {\bibfnamefont {I.~M.}\ \bibnamefont {Miron}}, \bibinfo {author}
  {\bibfnamefont {T.}~\bibnamefont {Jungwirth}}, \bibinfo {author}
  {\bibfnamefont {J.}~\bibnamefont {Sinova}}, \bibinfo {author} {\bibfnamefont
  {A.}~\bibnamefont {Thiaville}}, \bibinfo {author} {\bibfnamefont
  {K.}~\bibnamefont {Garello}}, \ and\ \bibinfo {author} {\bibfnamefont
  {P.}~\bibnamefont {Gambardella}},\ }\href {\doibase
  10.1103/RevModPhys.91.035004} {\bibfield  {journal} {\bibinfo  {journal}
  {Rev. Mod. Phys.}\ }\textbf {\bibinfo {volume} {91}},\ \bibinfo {pages}
  {035004} (\bibinfo {year} {2019})}\BibitemShut {NoStop}%
\bibitem [{\citenamefont {Shao}\ \emph {et~al.}(2021)\citenamefont {Shao},
  \citenamefont {Li}, \citenamefont {Liu}, \citenamefont {Yang}, \citenamefont
  {Fukami}, \citenamefont {Razavi}, \citenamefont {Wu}, \citenamefont {Wang},
  \citenamefont {Freimuth}, \citenamefont {Mokrousov}, \citenamefont {Stiles},
  \citenamefont {Emori}, \citenamefont {Hoffmann}, \citenamefont {Åkerman},
  \citenamefont {Roy}, \citenamefont {Wang}, \citenamefont {Yang},
  \citenamefont {Garello},\ and\ \citenamefont {Zhang}}]{Shao2021ieeetmag}%
  \BibitemOpen
  \bibfield  {author} {\bibinfo {author} {\bibfnamefont {Q.}~\bibnamefont
  {Shao}}, \bibinfo {author} {\bibfnamefont {P.}~\bibnamefont {Li}}, \bibinfo
  {author} {\bibfnamefont {L.}~\bibnamefont {Liu}}, \bibinfo {author}
  {\bibfnamefont {H.}~\bibnamefont {Yang}}, \bibinfo {author} {\bibfnamefont
  {S.}~\bibnamefont {Fukami}}, \bibinfo {author} {\bibfnamefont
  {A.}~\bibnamefont {Razavi}}, \bibinfo {author} {\bibfnamefont
  {H.}~\bibnamefont {Wu}}, \bibinfo {author} {\bibfnamefont {K.}~\bibnamefont
  {Wang}}, \bibinfo {author} {\bibfnamefont {F.}~\bibnamefont {Freimuth}},
  \bibinfo {author} {\bibfnamefont {Y.}~\bibnamefont {Mokrousov}}, \bibinfo
  {author} {\bibfnamefont {M.~D.}\ \bibnamefont {Stiles}}, \bibinfo {author}
  {\bibfnamefont {S.}~\bibnamefont {Emori}}, \bibinfo {author} {\bibfnamefont
  {A.}~\bibnamefont {Hoffmann}}, \bibinfo {author} {\bibfnamefont
  {J.}~\bibnamefont {Åkerman}}, \bibinfo {author} {\bibfnamefont
  {K.}~\bibnamefont {Roy}}, \bibinfo {author} {\bibfnamefont {J.-P.}\
  \bibnamefont {Wang}}, \bibinfo {author} {\bibfnamefont {S.-H.}\ \bibnamefont
  {Yang}}, \bibinfo {author} {\bibfnamefont {K.}~\bibnamefont {Garello}}, \
  and\ \bibinfo {author} {\bibfnamefont {W.}~\bibnamefont {Zhang}},\ }\href
  {\doibase 10.1109/TMAG.2021.3078583} {\bibfield  {journal} {\bibinfo
  {journal} {IEEE Transactions on Magnetics}\ }\textbf {\bibinfo {volume}
  {57}},\ \bibinfo {pages} {1} (\bibinfo {year} {2021})}\BibitemShut {NoStop}%
\bibitem [{\citenamefont {Dyakonov}\ and\ \citenamefont
  {Perel}(1971)}]{Dyakonov1971pl}%
  \BibitemOpen
  \bibfield  {author} {\bibinfo {author} {\bibfnamefont {M.~I.}\ \bibnamefont
  {Dyakonov}}\ and\ \bibinfo {author} {\bibfnamefont {V.~I.}\ \bibnamefont
  {Perel}},\ }\href {\doibase 10.1016/0375-9601(71)90196-4} {\bibfield
  {journal} {\bibinfo  {journal} {Phys. Lett. A}\ }\textbf {\bibinfo {volume}
  {35}},\ \bibinfo {pages} {459} (\bibinfo {year} {1971})}\BibitemShut
  {NoStop}%
\bibitem [{\citenamefont {Hirsch}(1999)}]{Hirsch1999prl}%
  \BibitemOpen
  \bibfield  {author} {\bibinfo {author} {\bibfnamefont {J.~E.}\ \bibnamefont
  {Hirsch}},\ }\href@noop {} {\bibfield  {journal} {\bibinfo  {journal} {Phys.
  Rev. Lett.}\ }\textbf {\bibinfo {volume} {83}},\ \bibinfo {pages} {1834}
  (\bibinfo {year} {1999})}\BibitemShut {NoStop}%
\bibitem [{\citenamefont {Sinova}\ \emph {et~al.}(2015)\citenamefont {Sinova},
  \citenamefont {Valenzuela}, \citenamefont {Wunderlich}, \citenamefont
  {Back},\ and\ \citenamefont {Jungwirth}}]{Sinova2015RevModPhys}%
  \BibitemOpen
  \bibfield  {author} {\bibinfo {author} {\bibfnamefont {J.}~\bibnamefont
  {Sinova}}, \bibinfo {author} {\bibfnamefont {S.~O.}\ \bibnamefont
  {Valenzuela}}, \bibinfo {author} {\bibfnamefont {J.}~\bibnamefont
  {Wunderlich}}, \bibinfo {author} {\bibfnamefont {C.~H.}\ \bibnamefont
  {Back}}, \ and\ \bibinfo {author} {\bibfnamefont {T.}~\bibnamefont
  {Jungwirth}},\ }\href {\doibase 10.1103/RevModPhys.87.1213} {\bibfield
  {journal} {\bibinfo  {journal} {Rev. Mod. Phys.}\ }\textbf {\bibinfo {volume}
  {87}},\ \bibinfo {pages} {1213} (\bibinfo {year} {2015})}\BibitemShut
  {NoStop}%
\bibitem [{\citenamefont {Demidov}\ \emph {et~al.}(2012)\citenamefont
  {Demidov}, \citenamefont {Urazhdin}, \citenamefont {Ulrichs}, \citenamefont
  {Tiberkevich}, \citenamefont {Slavin}, \citenamefont {Baither}, \citenamefont
  {Schmitz},\ and\ \citenamefont {Demokritov}}]{Demidov2012b}%
  \BibitemOpen
  \bibfield  {author} {\bibinfo {author} {\bibfnamefont {V.~E.}\ \bibnamefont
  {Demidov}}, \bibinfo {author} {\bibfnamefont {S.}~\bibnamefont {Urazhdin}},
  \bibinfo {author} {\bibfnamefont {H.}~\bibnamefont {Ulrichs}}, \bibinfo
  {author} {\bibfnamefont {V.}~\bibnamefont {Tiberkevich}}, \bibinfo {author}
  {\bibfnamefont {A.}~\bibnamefont {Slavin}}, \bibinfo {author} {\bibfnamefont
  {D.}~\bibnamefont {Baither}}, \bibinfo {author} {\bibfnamefont
  {G.}~\bibnamefont {Schmitz}}, \ and\ \bibinfo {author} {\bibfnamefont
  {S.~O.}\ \bibnamefont {Demokritov}},\ }\href {\doibase 10.1038/nmat3459}
  {\bibfield  {journal} {\bibinfo  {journal} {Nat. Mater.}\ }\textbf {\bibinfo
  {volume} {11}},\ \bibinfo {pages} {1028} (\bibinfo {year}
  {2012})}\BibitemShut {NoStop}%
\bibitem [{\citenamefont {Liu}\ \emph {et~al.}(2012{\natexlab{a}})\citenamefont
  {Liu}, \citenamefont {Pai}, \citenamefont {Ralph},\ and\ \citenamefont
  {Buhrman}}]{liu2012prl}%
  \BibitemOpen
  \bibfield  {author} {\bibinfo {author} {\bibfnamefont {L.}~\bibnamefont
  {Liu}}, \bibinfo {author} {\bibfnamefont {C.-F.}\ \bibnamefont {Pai}},
  \bibinfo {author} {\bibfnamefont {D.~C.}\ \bibnamefont {Ralph}}, \ and\
  \bibinfo {author} {\bibfnamefont {R.~A.}\ \bibnamefont {Buhrman}},\ }\href
  {\doibase 10.1103/PhysRevLett.109.186602} {\bibfield  {journal} {\bibinfo
  {journal} {Phys. Rev. Lett.}\ }\textbf {\bibinfo {volume} {109}},\ \bibinfo
  {pages} {186602} (\bibinfo {year} {2012}{\natexlab{a}})}\BibitemShut
  {NoStop}%
\bibitem [{\citenamefont {Demidov}\ \emph {et~al.}(2014)\citenamefont
  {Demidov}, \citenamefont {Urazhdin}, \citenamefont {Zholud}, \citenamefont
  {Sadovnikov},\ and\ \citenamefont {Demokritov}}]{Demidov2014apl}%
  \BibitemOpen
  \bibfield  {author} {\bibinfo {author} {\bibfnamefont {V.~E.}\ \bibnamefont
  {Demidov}}, \bibinfo {author} {\bibfnamefont {S.}~\bibnamefont {Urazhdin}},
  \bibinfo {author} {\bibfnamefont {A.}~\bibnamefont {Zholud}}, \bibinfo
  {author} {\bibfnamefont {A.~V.}\ \bibnamefont {Sadovnikov}}, \ and\ \bibinfo
  {author} {\bibfnamefont {S.~O.}\ \bibnamefont {Demokritov}},\ }\href
  {\doibase 10.1063/1.4901027} {\bibfield  {journal} {\bibinfo  {journal}
  {Appl. Phys. Lett.}\ }\textbf {\bibinfo {volume} {105}},\ \bibinfo {pages}
  {172410} (\bibinfo {year} {2014})}\BibitemShut {NoStop}%
\bibitem [{\citenamefont {Ranjbar}\ \emph {et~al.}(2014)\citenamefont
  {Ranjbar}, \citenamefont {D{\"{u}}rrenfeld}, \citenamefont {Haidar},
  \citenamefont {Iacocca}, \citenamefont {Balinskiy}, \citenamefont {Le},
  \citenamefont {Fazlali}, \citenamefont {Houshang}, \citenamefont {Awad},
  \citenamefont {Dumas},\ and\ \citenamefont {{\AA}kerman}}]{Ranjbar2014}%
  \BibitemOpen
  \bibfield  {author} {\bibinfo {author} {\bibfnamefont {M.}~\bibnamefont
  {Ranjbar}}, \bibinfo {author} {\bibfnamefont {P.}~\bibnamefont
  {D{\"{u}}rrenfeld}}, \bibinfo {author} {\bibfnamefont {M.}~\bibnamefont
  {Haidar}}, \bibinfo {author} {\bibfnamefont {E.}~\bibnamefont {Iacocca}},
  \bibinfo {author} {\bibfnamefont {M.}~\bibnamefont {Balinskiy}}, \bibinfo
  {author} {\bibfnamefont {T.~Q.}\ \bibnamefont {Le}}, \bibinfo {author}
  {\bibfnamefont {M.}~\bibnamefont {Fazlali}}, \bibinfo {author} {\bibfnamefont
  {A.}~\bibnamefont {Houshang}}, \bibinfo {author} {\bibfnamefont {A.~A.}\
  \bibnamefont {Awad}}, \bibinfo {author} {\bibfnamefont {R.~K.}\ \bibnamefont
  {Dumas}}, \ and\ \bibinfo {author} {\bibfnamefont {J.}~\bibnamefont
  {{\AA}kerman}},\ }\href {\doibase 10.1109/LMAG.2014.2375155} {\bibfield
  {journal} {\bibinfo  {journal} {IEEE Magn. Lett.}\ }\textbf {\bibinfo
  {volume} {5}},\ \bibinfo {pages} {3000504} (\bibinfo {year}
  {2014})}\BibitemShut {NoStop}%
\bibitem [{\citenamefont {Chen}\ \emph {et~al.}(2016)\citenamefont {Chen},
  \citenamefont {Dumas}, \citenamefont {Eklund}, \citenamefont {Muduli},
  \citenamefont {Houshang}, \citenamefont {Awad}, \citenamefont
  {D\"{u}rrenfeld}, \citenamefont {Malm}, \citenamefont {Rusu},\ and\
  \citenamefont {{\AA}kerman}}]{Chen2016procieee}%
  \BibitemOpen
  \bibfield  {author} {\bibinfo {author} {\bibfnamefont {T.}~\bibnamefont
  {Chen}}, \bibinfo {author} {\bibfnamefont {R.~K.}\ \bibnamefont {Dumas}},
  \bibinfo {author} {\bibfnamefont {A.}~\bibnamefont {Eklund}}, \bibinfo
  {author} {\bibfnamefont {P.~K.}\ \bibnamefont {Muduli}}, \bibinfo {author}
  {\bibfnamefont {A.}~\bibnamefont {Houshang}}, \bibinfo {author}
  {\bibfnamefont {A.~A.}\ \bibnamefont {Awad}}, \bibinfo {author} {\bibnamefont
  {D\"{u}rrenfeld}}, \bibinfo {author} {\bibfnamefont {B.~G.}\ \bibnamefont
  {Malm}}, \bibinfo {author} {\bibfnamefont {A.}~\bibnamefont {Rusu}}, \ and\
  \bibinfo {author} {\bibfnamefont {J.}~\bibnamefont {{\AA}kerman}},\ }\href
  {\doibase 10.1109/JPROC.2016.2554518} {\bibfield  {journal} {\bibinfo
  {journal} {P. IEEE}\ }\textbf {\bibinfo {volume} {104}},\ \bibinfo {pages}
  {1919} (\bibinfo {year} {2016})}\BibitemShut {NoStop}%
\bibitem [{\citenamefont {Zahedinejad}\ \emph {et~al.}(2018)\citenamefont
  {Zahedinejad}, \citenamefont {Mazraati}, \citenamefont {Fulara},
  \citenamefont {Yue}, \citenamefont {Jiang}, \citenamefont {Awad},\ and\
  \citenamefont {{\AA}kerman}}]{Zahedinejad2018apl}%
  \BibitemOpen
  \bibfield  {author} {\bibinfo {author} {\bibfnamefont {M.}~\bibnamefont
  {Zahedinejad}}, \bibinfo {author} {\bibfnamefont {H.}~\bibnamefont
  {Mazraati}}, \bibinfo {author} {\bibfnamefont {H.}~\bibnamefont {Fulara}},
  \bibinfo {author} {\bibfnamefont {J.}~\bibnamefont {Yue}}, \bibinfo {author}
  {\bibfnamefont {S.}~\bibnamefont {Jiang}}, \bibinfo {author} {\bibfnamefont
  {A.~A.}\ \bibnamefont {Awad}}, \ and\ \bibinfo {author} {\bibfnamefont
  {J.}~\bibnamefont {{\AA}kerman}},\ }\href {\doibase 10.1063/1.5022049}
  {\bibfield  {journal} {\bibinfo  {journal} {Appl. Phys. Lett.}\ }\textbf
  {\bibinfo {volume} {112}},\ \bibinfo {pages} {132404} (\bibinfo {year}
  {2018})}\BibitemShut {NoStop}%
\bibitem [{\citenamefont {Haidar}\ \emph {et~al.}(2019)\citenamefont {Haidar},
  \citenamefont {Awad}, \citenamefont {Dvornik}, \citenamefont {Khymyn},
  \citenamefont {Houshang},\ and\ \citenamefont
  {{\AA}kerman}}]{Haidar2019natcomm}%
  \BibitemOpen
  \bibfield  {author} {\bibinfo {author} {\bibfnamefont {M.}~\bibnamefont
  {Haidar}}, \bibinfo {author} {\bibfnamefont {A.~A.}\ \bibnamefont {Awad}},
  \bibinfo {author} {\bibfnamefont {M.}~\bibnamefont {Dvornik}}, \bibinfo
  {author} {\bibfnamefont {R.}~\bibnamefont {Khymyn}}, \bibinfo {author}
  {\bibfnamefont {A.}~\bibnamefont {Houshang}}, \ and\ \bibinfo {author}
  {\bibfnamefont {J.}~\bibnamefont {{\AA}kerman}},\ }\href {\doibase
  10.1038/s41467-019-10120-4} {\bibfield  {journal} {\bibinfo  {journal}
  {Nature Communications}\ }\textbf {\bibinfo {volume} {10}},\ \bibinfo {pages}
  {2362} (\bibinfo {year} {2019})}\BibitemShut {NoStop}%
\bibitem [{\citenamefont {D{\"{u}}rrenfeld}\ \emph {et~al.}(2017)\citenamefont
  {D{\"{u}}rrenfeld}, \citenamefont {Awad}, \citenamefont {Houshang},
  \citenamefont {Dumas},\ and\ \citenamefont
  {\AA~kerman}}]{durrenfeld2017nanoscale}%
  \BibitemOpen
  \bibfield  {author} {\bibinfo {author} {\bibfnamefont {P.}~\bibnamefont
  {D{\"{u}}rrenfeld}}, \bibinfo {author} {\bibfnamefont {A.~A.}\ \bibnamefont
  {Awad}}, \bibinfo {author} {\bibfnamefont {A.}~\bibnamefont {Houshang}},
  \bibinfo {author} {\bibfnamefont {R.~K.}\ \bibnamefont {Dumas}}, \ and\
  \bibinfo {author} {\bibfnamefont {J.}~\bibnamefont {\AA~kerman}},\ }\href
  {\doibase 10.1039/C6NR07903B} {\bibfield  {journal} {\bibinfo  {journal}
  {Nanoscale}\ }\textbf {\bibinfo {volume} {9}},\ \bibinfo {pages} {1285}
  (\bibinfo {year} {2017})}\BibitemShut {NoStop}%
\bibitem [{\citenamefont {Duan}\ \emph {et~al.}(2014)\citenamefont {Duan},
  \citenamefont {Smith}, \citenamefont {Yang}, \citenamefont {Youngblood},
  \citenamefont {Lindner}, \citenamefont {Demidov}, \citenamefont
  {Demokritov},\ and\ \citenamefont {Krivorotov}}]{Duan2014b}%
  \BibitemOpen
  \bibfield  {author} {\bibinfo {author} {\bibfnamefont {Z.}~\bibnamefont
  {Duan}}, \bibinfo {author} {\bibfnamefont {A.}~\bibnamefont {Smith}},
  \bibinfo {author} {\bibfnamefont {L.}~\bibnamefont {Yang}}, \bibinfo {author}
  {\bibfnamefont {B.}~\bibnamefont {Youngblood}}, \bibinfo {author}
  {\bibfnamefont {J.}~\bibnamefont {Lindner}}, \bibinfo {author} {\bibfnamefont
  {V.~E.}\ \bibnamefont {Demidov}}, \bibinfo {author} {\bibfnamefont {S.~O.}\
  \bibnamefont {Demokritov}}, \ and\ \bibinfo {author} {\bibfnamefont {I.~N.}\
  \bibnamefont {Krivorotov}},\ }\href {\doibase 10.1038/ncomms6616} {\bibfield
  {journal} {\bibinfo  {journal} {Nat. Commun.}\ }\textbf {\bibinfo {volume}
  {5}},\ \bibinfo {pages} {5616} (\bibinfo {year} {2014})}\BibitemShut
  {NoStop}%
\bibitem [{\citenamefont {Liu}\ \emph {et~al.}(2012{\natexlab{b}})\citenamefont
  {Liu}, \citenamefont {Pai}, \citenamefont {Li}, \citenamefont {Tseng},
  \citenamefont {Ralph},\ and\ \citenamefont {Buhrman}}]{liu2012science}%
  \BibitemOpen
  \bibfield  {author} {\bibinfo {author} {\bibfnamefont {L.}~\bibnamefont
  {Liu}}, \bibinfo {author} {\bibfnamefont {C.-F.}\ \bibnamefont {Pai}},
  \bibinfo {author} {\bibfnamefont {Y.}~\bibnamefont {Li}}, \bibinfo {author}
  {\bibfnamefont {H.~W.}\ \bibnamefont {Tseng}}, \bibinfo {author}
  {\bibfnamefont {D.~C.}\ \bibnamefont {Ralph}}, \ and\ \bibinfo {author}
  {\bibfnamefont {R.~A.}\ \bibnamefont {Buhrman}},\ }\href {\doibase
  10.1126/science.1218197} {\bibfield  {journal} {\bibinfo  {journal}
  {Science}\ }\textbf {\bibinfo {volume} {336}},\ \bibinfo {pages} {555}
  (\bibinfo {year} {2012}{\natexlab{b}})}\BibitemShut {NoStop}%
\bibitem [{\citenamefont {Miron}\ \emph {et~al.}(2011)\citenamefont {Miron},
  \citenamefont {Garello}, \citenamefont {Gaudin}, \citenamefont {Zermatten},
  \citenamefont {Costache}, \citenamefont {Auffret}, \citenamefont {Bandiera},
  \citenamefont {Rodmacq}, \citenamefont {Schuhl},\ and\ \citenamefont
  {Gambardella}}]{Miron2011nat}%
  \BibitemOpen
  \bibfield  {author} {\bibinfo {author} {\bibfnamefont {I.~M.}\ \bibnamefont
  {Miron}}, \bibinfo {author} {\bibfnamefont {K.}~\bibnamefont {Garello}},
  \bibinfo {author} {\bibfnamefont {G.}~\bibnamefont {Gaudin}}, \bibinfo
  {author} {\bibfnamefont {P.-J.}\ \bibnamefont {Zermatten}}, \bibinfo {author}
  {\bibfnamefont {M.~V.}\ \bibnamefont {Costache}}, \bibinfo {author}
  {\bibfnamefont {S.}~\bibnamefont {Auffret}}, \bibinfo {author} {\bibfnamefont
  {S.}~\bibnamefont {Bandiera}}, \bibinfo {author} {\bibfnamefont
  {B.}~\bibnamefont {Rodmacq}}, \bibinfo {author} {\bibfnamefont
  {A.}~\bibnamefont {Schuhl}}, \ and\ \bibinfo {author} {\bibfnamefont
  {P.}~\bibnamefont {Gambardella}},\ }\href {\doibase 10.1038/nature10309}
  {\bibfield  {journal} {\bibinfo  {journal} {Nature}\ }\textbf {\bibinfo
  {volume} {476}},\ \bibinfo {pages} {189} (\bibinfo {year}
  {2011})}\BibitemShut {NoStop}%
\bibitem [{\citenamefont {Xing}\ \emph {et~al.}(2017)\citenamefont {Xing},
  \citenamefont {Pong}, \citenamefont {\AA{}kerman},\ and\ \citenamefont
  {Zhou}}]{Xing2017prappl}%
  \BibitemOpen
  \bibfield  {author} {\bibinfo {author} {\bibfnamefont {X.}~\bibnamefont
  {Xing}}, \bibinfo {author} {\bibfnamefont {P.~W.~T.}\ \bibnamefont {Pong}},
  \bibinfo {author} {\bibfnamefont {J.}~\bibnamefont {\AA{}kerman}}, \ and\
  \bibinfo {author} {\bibfnamefont {Y.}~\bibnamefont {Zhou}},\ }\href {\doibase
  10.1103/PhysRevApplied.7.054016} {\bibfield  {journal} {\bibinfo  {journal}
  {Phys. Rev. Applied}\ }\textbf {\bibinfo {volume} {7}},\ \bibinfo {pages}
  {054016} (\bibinfo {year} {2017})}\BibitemShut {NoStop}%
\bibitem [{\citenamefont {Demidov}\ \emph {et~al.}(2020)\citenamefont
  {Demidov}, \citenamefont {Urazhdin}, \citenamefont {Anane}, \citenamefont
  {Cros},\ and\ \citenamefont {Demokritov}}]{Demidov2020jap}%
  \BibitemOpen
  \bibfield  {author} {\bibinfo {author} {\bibfnamefont {V.~E.}\ \bibnamefont
  {Demidov}}, \bibinfo {author} {\bibfnamefont {S.}~\bibnamefont {Urazhdin}},
  \bibinfo {author} {\bibfnamefont {A.}~\bibnamefont {Anane}}, \bibinfo
  {author} {\bibfnamefont {V.}~\bibnamefont {Cros}}, \ and\ \bibinfo {author}
  {\bibfnamefont {S.~O.}\ \bibnamefont {Demokritov}},\ }\href {\doibase
  10.1063/5.0007095} {\bibfield  {journal} {\bibinfo  {journal} {Journal of
  Applied Physics}\ }\textbf {\bibinfo {volume} {127}},\ \bibinfo {pages}
  {170901} (\bibinfo {year} {2020})},\ \Eprint
  {http://arxiv.org/abs/https://doi.org/10.1063/5.0007095}
  {https://doi.org/10.1063/5.0007095} \BibitemShut {NoStop}%
\bibitem [{\citenamefont {Dieny}\ \emph {et~al.}(2020)\citenamefont {Dieny},
  \citenamefont {Prejbeanu}, \citenamefont {Garello}, \citenamefont
  {Gambardella}, \citenamefont {Freitas}, \citenamefont {Lehndorff},
  \citenamefont {Raberg}, \citenamefont {Ebels}, \citenamefont {Demokritov},
  \citenamefont {Akerman}, \citenamefont {Deac}, \citenamefont {Pirro},
  \citenamefont {Adelmann}, \citenamefont {Anane}, \citenamefont {Chumak},
  \citenamefont {Hirohata}, \citenamefont {Mangin}, \citenamefont {Valenzuela},
  \citenamefont {Onba{\c{s}}l{\i}}, \citenamefont {d'Aquino}, \citenamefont
  {Prenat}, \citenamefont {Finocchio}, \citenamefont {Lopez-Diaz},
  \citenamefont {Chantrell}, \citenamefont {Chubykalo-Fesenko},\ and\
  \citenamefont {Bortolotti}}]{Dieny2020natelec}%
  \BibitemOpen
  \bibfield  {author} {\bibinfo {author} {\bibfnamefont {B.}~\bibnamefont
  {Dieny}}, \bibinfo {author} {\bibfnamefont {I.~L.}\ \bibnamefont
  {Prejbeanu}}, \bibinfo {author} {\bibfnamefont {K.}~\bibnamefont {Garello}},
  \bibinfo {author} {\bibfnamefont {P.}~\bibnamefont {Gambardella}}, \bibinfo
  {author} {\bibfnamefont {P.}~\bibnamefont {Freitas}}, \bibinfo {author}
  {\bibfnamefont {R.}~\bibnamefont {Lehndorff}}, \bibinfo {author}
  {\bibfnamefont {W.}~\bibnamefont {Raberg}}, \bibinfo {author} {\bibfnamefont
  {U.}~\bibnamefont {Ebels}}, \bibinfo {author} {\bibfnamefont {S.~O.}\
  \bibnamefont {Demokritov}}, \bibinfo {author} {\bibfnamefont
  {J.}~\bibnamefont {Akerman}}, \bibinfo {author} {\bibfnamefont
  {A.}~\bibnamefont {Deac}}, \bibinfo {author} {\bibfnamefont {P.}~\bibnamefont
  {Pirro}}, \bibinfo {author} {\bibfnamefont {C.}~\bibnamefont {Adelmann}},
  \bibinfo {author} {\bibfnamefont {A.}~\bibnamefont {Anane}}, \bibinfo
  {author} {\bibfnamefont {A.~V.}\ \bibnamefont {Chumak}}, \bibinfo {author}
  {\bibfnamefont {A.}~\bibnamefont {Hirohata}}, \bibinfo {author}
  {\bibfnamefont {S.}~\bibnamefont {Mangin}}, \bibinfo {author} {\bibfnamefont
  {S.~O.}\ \bibnamefont {Valenzuela}}, \bibinfo {author} {\bibfnamefont
  {M.~C.}\ \bibnamefont {Onba{\c{s}}l{\i}}}, \bibinfo {author} {\bibfnamefont
  {M.}~\bibnamefont {d'Aquino}}, \bibinfo {author} {\bibfnamefont
  {G.}~\bibnamefont {Prenat}}, \bibinfo {author} {\bibfnamefont
  {G.}~\bibnamefont {Finocchio}}, \bibinfo {author} {\bibfnamefont
  {L.}~\bibnamefont {Lopez-Diaz}}, \bibinfo {author} {\bibfnamefont
  {R.}~\bibnamefont {Chantrell}}, \bibinfo {author} {\bibfnamefont
  {O.}~\bibnamefont {Chubykalo-Fesenko}}, \ and\ \bibinfo {author}
  {\bibfnamefont {P.}~\bibnamefont {Bortolotti}},\ }\href {\doibase
  10.1038/s41928-020-0461-5} {\bibfield  {journal} {\bibinfo  {journal} {Nature
  Electronics}\ }\textbf {\bibinfo {volume} {3}},\ \bibinfo {pages} {446}
  (\bibinfo {year} {2020})}\BibitemShut {NoStop}%
\bibitem [{\citenamefont {Awad}\ \emph {et~al.}(2020)\citenamefont {Awad},
  \citenamefont {Houshang}, \citenamefont {Zahedinejad}, \citenamefont
  {Khymyn},\ and\ \citenamefont {Åkerman}}]{Awad2020apl}%
  \BibitemOpen
  \bibfield  {author} {\bibinfo {author} {\bibfnamefont {A.~A.}\ \bibnamefont
  {Awad}}, \bibinfo {author} {\bibfnamefont {A.}~\bibnamefont {Houshang}},
  \bibinfo {author} {\bibfnamefont {M.}~\bibnamefont {Zahedinejad}}, \bibinfo
  {author} {\bibfnamefont {R.}~\bibnamefont {Khymyn}}, \ and\ \bibinfo {author}
  {\bibfnamefont {J.}~\bibnamefont {Åkerman}},\ }\href {\doibase
  10.1063/5.0007254} {\bibfield  {journal} {\bibinfo  {journal} {Applied
  Physics Letters}\ }\textbf {\bibinfo {volume} {116}},\ \bibinfo {pages}
  {232401} (\bibinfo {year} {2020})},\ \Eprint
  {http://arxiv.org/abs/https://doi.org/10.1063/5.0007254}
  {https://doi.org/10.1063/5.0007254} \BibitemShut {NoStop}%
\bibitem [{\citenamefont {Zahedinejad}\ \emph {et~al.}(2017)\citenamefont
  {Zahedinejad}, \citenamefont {Awad}, \citenamefont {Dürrenfeld},
  \citenamefont {Houshang}, \citenamefont {Yin}, \citenamefont {Muduli},\ and\
  \citenamefont {Åkerman}}]{Zahedinejad2017ieeeml}%
  \BibitemOpen
  \bibfield  {author} {\bibinfo {author} {\bibfnamefont {M.}~\bibnamefont
  {Zahedinejad}}, \bibinfo {author} {\bibfnamefont {A.~A.}\ \bibnamefont
  {Awad}}, \bibinfo {author} {\bibfnamefont {P.}~\bibnamefont {Dürrenfeld}},
  \bibinfo {author} {\bibfnamefont {A.}~\bibnamefont {Houshang}}, \bibinfo
  {author} {\bibfnamefont {Y.}~\bibnamefont {Yin}}, \bibinfo {author}
  {\bibfnamefont {P.~K.}\ \bibnamefont {Muduli}}, \ and\ \bibinfo {author}
  {\bibfnamefont {J.}~\bibnamefont {Åkerman}},\ }\href {\doibase
  10.1109/LMAG.2017.2671453} {\bibfield  {journal} {\bibinfo  {journal} {IEEE
  Magnetics Letters}\ }\textbf {\bibinfo {volume} {8}},\ \bibinfo {pages} {1}
  (\bibinfo {year} {2017})}\BibitemShut {NoStop}%
\bibitem [{\citenamefont {Dvornik}\ \emph {et~al.}(2018)\citenamefont
  {Dvornik}, \citenamefont {Awad},\ and\ \citenamefont
  {\AA{}kerman}}]{Dvornik2018prappl}%
  \BibitemOpen
  \bibfield  {author} {\bibinfo {author} {\bibfnamefont {M.}~\bibnamefont
  {Dvornik}}, \bibinfo {author} {\bibfnamefont {A.~A.}\ \bibnamefont {Awad}}, \
  and\ \bibinfo {author} {\bibfnamefont {J.}~\bibnamefont {\AA{}kerman}},\
  }\href {\doibase 10.1103/PhysRevApplied.9.014017} {\bibfield  {journal}
  {\bibinfo  {journal} {Phys. Rev. Applied}\ }\textbf {\bibinfo {volume} {9}},\
  \bibinfo {pages} {014017} (\bibinfo {year} {2018})}\BibitemShut {NoStop}%
\bibitem [{\citenamefont {Spicer}\ \emph {et~al.}(2018)\citenamefont {Spicer},
  \citenamefont {Keatley}, \citenamefont {Loughran}, \citenamefont {Dvornik},
  \citenamefont {Awad}, \citenamefont {D\"urrenfeld}, \citenamefont {Houshang},
  \citenamefont {Ranjbar}, \citenamefont {\AA{}kerman}, \citenamefont
  {Kruglyak},\ and\ \citenamefont {Hicken}}]{Spicer2018prb}%
  \BibitemOpen
  \bibfield  {author} {\bibinfo {author} {\bibfnamefont {T.~M.}\ \bibnamefont
  {Spicer}}, \bibinfo {author} {\bibfnamefont {P.~S.}\ \bibnamefont {Keatley}},
  \bibinfo {author} {\bibfnamefont {T.~H.~J.}\ \bibnamefont {Loughran}},
  \bibinfo {author} {\bibfnamefont {M.}~\bibnamefont {Dvornik}}, \bibinfo
  {author} {\bibfnamefont {A.~A.}\ \bibnamefont {Awad}}, \bibinfo {author}
  {\bibfnamefont {P.}~\bibnamefont {D\"urrenfeld}}, \bibinfo {author}
  {\bibfnamefont {A.}~\bibnamefont {Houshang}}, \bibinfo {author}
  {\bibfnamefont {M.}~\bibnamefont {Ranjbar}}, \bibinfo {author} {\bibfnamefont
  {J.}~\bibnamefont {\AA{}kerman}}, \bibinfo {author} {\bibfnamefont {V.~V.}\
  \bibnamefont {Kruglyak}}, \ and\ \bibinfo {author} {\bibfnamefont {R.~J.}\
  \bibnamefont {Hicken}},\ }\href {\doibase 10.1103/PhysRevB.98.214438}
  {\bibfield  {journal} {\bibinfo  {journal} {Phys. Rev. B}\ }\textbf {\bibinfo
  {volume} {98}},\ \bibinfo {pages} {214438} (\bibinfo {year}
  {2018})}\BibitemShut {NoStop}%
\bibitem [{\citenamefont {Mazraati}\ \emph
  {et~al.}(2018{\natexlab{a}})\citenamefont {Mazraati}, \citenamefont
  {Etesami}, \citenamefont {Banuazizi}, \citenamefont {Chung}, \citenamefont
  {Houshang}, \citenamefont {Awad}, \citenamefont {Dvornik},\ and\
  \citenamefont {\AA{}kerman}}]{mazraati2018prappl}%
  \BibitemOpen
  \bibfield  {author} {\bibinfo {author} {\bibfnamefont {H.}~\bibnamefont
  {Mazraati}}, \bibinfo {author} {\bibfnamefont {S.~R.}\ \bibnamefont
  {Etesami}}, \bibinfo {author} {\bibfnamefont {S.~A.~H.}\ \bibnamefont
  {Banuazizi}}, \bibinfo {author} {\bibfnamefont {S.}~\bibnamefont {Chung}},
  \bibinfo {author} {\bibfnamefont {A.}~\bibnamefont {Houshang}}, \bibinfo
  {author} {\bibfnamefont {A.~A.}\ \bibnamefont {Awad}}, \bibinfo {author}
  {\bibfnamefont {M.}~\bibnamefont {Dvornik}}, \ and\ \bibinfo {author}
  {\bibfnamefont {J.}~\bibnamefont {\AA{}kerman}},\ }\href {\doibase
  10.1103/PhysRevApplied.10.054017} {\bibfield  {journal} {\bibinfo  {journal}
  {Phys. Rev. Applied}\ }\textbf {\bibinfo {volume} {10}},\ \bibinfo {pages}
  {054017} (\bibinfo {year} {2018}{\natexlab{a}})}\BibitemShut {NoStop}%
\bibitem [{\citenamefont {Hache}\ \emph {et~al.}(2019)\citenamefont {Hache},
  \citenamefont {Weinhold}, \citenamefont {Schultheiss}, \citenamefont
  {Stigloher}, \citenamefont {Vilsmeier}, \citenamefont {Back}, \citenamefont
  {Arekapudi}, \citenamefont {Hellwig}, \citenamefont {Fassbender},\ and\
  \citenamefont {Schultheiss}}]{Hache2019apl}%
  \BibitemOpen
  \bibfield  {author} {\bibinfo {author} {\bibfnamefont {T.}~\bibnamefont
  {Hache}}, \bibinfo {author} {\bibfnamefont {T.}~\bibnamefont {Weinhold}},
  \bibinfo {author} {\bibfnamefont {K.}~\bibnamefont {Schultheiss}}, \bibinfo
  {author} {\bibfnamefont {J.}~\bibnamefont {Stigloher}}, \bibinfo {author}
  {\bibfnamefont {F.}~\bibnamefont {Vilsmeier}}, \bibinfo {author}
  {\bibfnamefont {C.}~\bibnamefont {Back}}, \bibinfo {author} {\bibfnamefont
  {S.~S. P.~K.}\ \bibnamefont {Arekapudi}}, \bibinfo {author} {\bibfnamefont
  {O.}~\bibnamefont {Hellwig}}, \bibinfo {author} {\bibfnamefont
  {J.}~\bibnamefont {Fassbender}}, \ and\ \bibinfo {author} {\bibfnamefont
  {H.}~\bibnamefont {Schultheiss}},\ }\href {\doibase 10.1063/1.5082692}
  {\bibfield  {journal} {\bibinfo  {journal} {Applied Physics Letters}\
  }\textbf {\bibinfo {volume} {114}},\ \bibinfo {pages} {102403} (\bibinfo
  {year} {2019})},\ \Eprint
  {http://arxiv.org/abs/https://doi.org/10.1063/1.5082692}
  {https://doi.org/10.1063/1.5082692} \BibitemShut {NoStop}%
\bibitem [{\citenamefont {Smith}\ \emph {et~al.}(2020)\citenamefont {Smith},
  \citenamefont {Sobotkiewich}, \citenamefont {Khan}, \citenamefont {Montoya},
  \citenamefont {Yang}, \citenamefont {Duan}, \citenamefont {Schneider},
  \citenamefont {Lenz}, \citenamefont {Lindner}, \citenamefont {An},
  \citenamefont {Li},\ and\ \citenamefont {Krivorotov}}]{Smith2020prb}%
  \BibitemOpen
  \bibfield  {author} {\bibinfo {author} {\bibfnamefont {A.}~\bibnamefont
  {Smith}}, \bibinfo {author} {\bibfnamefont {K.}~\bibnamefont {Sobotkiewich}},
  \bibinfo {author} {\bibfnamefont {A.}~\bibnamefont {Khan}}, \bibinfo {author}
  {\bibfnamefont {E.~A.}\ \bibnamefont {Montoya}}, \bibinfo {author}
  {\bibfnamefont {L.}~\bibnamefont {Yang}}, \bibinfo {author} {\bibfnamefont
  {Z.}~\bibnamefont {Duan}}, \bibinfo {author} {\bibfnamefont {T.}~\bibnamefont
  {Schneider}}, \bibinfo {author} {\bibfnamefont {K.}~\bibnamefont {Lenz}},
  \bibinfo {author} {\bibfnamefont {J.}~\bibnamefont {Lindner}}, \bibinfo
  {author} {\bibfnamefont {K.}~\bibnamefont {An}}, \bibinfo {author}
  {\bibfnamefont {X.}~\bibnamefont {Li}}, \ and\ \bibinfo {author}
  {\bibfnamefont {I.~N.}\ \bibnamefont {Krivorotov}},\ }\href {\doibase
  10.1103/PhysRevB.102.054422} {\bibfield  {journal} {\bibinfo  {journal}
  {Phys. Rev. B}\ }\textbf {\bibinfo {volume} {102}},\ \bibinfo {pages}
  {054422} (\bibinfo {year} {2020})}\BibitemShut {NoStop}%
\bibitem [{\citenamefont {Divinskiy}\ \emph {et~al.}(2017)\citenamefont
  {Divinskiy}, \citenamefont {Demidov}, \citenamefont {Urazhdin}, \citenamefont
  {Freeman}, \citenamefont {Rinkevich},\ and\ \citenamefont
  {Demokritov}}]{divinskiy2017advm}%
  \BibitemOpen
  \bibfield  {author} {\bibinfo {author} {\bibfnamefont {B.}~\bibnamefont
  {Divinskiy}}, \bibinfo {author} {\bibfnamefont {V.~E.}\ \bibnamefont
  {Demidov}}, \bibinfo {author} {\bibfnamefont {S.}~\bibnamefont {Urazhdin}},
  \bibinfo {author} {\bibfnamefont {R.}~\bibnamefont {Freeman}}, \bibinfo
  {author} {\bibfnamefont {A.~B.}\ \bibnamefont {Rinkevich}}, \ and\ \bibinfo
  {author} {\bibfnamefont {S.~O.}\ \bibnamefont {Demokritov}},\ }\href
  {https://doi.org/10.1002/adma.201802837} {\bibfield  {journal} {\bibinfo
  {journal} {Advanced Materials}\ }\textbf {\bibinfo {volume} {30}},\ \bibinfo
  {pages} {1802837} (\bibinfo {year} {2017})}\BibitemShut {NoStop}%
\bibitem [{\citenamefont {Fulara}\ \emph {et~al.}(2019)\citenamefont {Fulara},
  \citenamefont {Zahedinejad}, \citenamefont {Khymyn}, \citenamefont {Awad},
  \citenamefont {Muralidhar}, \citenamefont {Dvornik},\ and\ \citenamefont {{\r
  A}kerman}}]{Fulara2019SciAdv}%
  \BibitemOpen
  \bibfield  {author} {\bibinfo {author} {\bibfnamefont {H.}~\bibnamefont
  {Fulara}}, \bibinfo {author} {\bibfnamefont {M.}~\bibnamefont {Zahedinejad}},
  \bibinfo {author} {\bibfnamefont {R.}~\bibnamefont {Khymyn}}, \bibinfo
  {author} {\bibfnamefont {A.~A.}\ \bibnamefont {Awad}}, \bibinfo {author}
  {\bibfnamefont {S.}~\bibnamefont {Muralidhar}}, \bibinfo {author}
  {\bibfnamefont {M.}~\bibnamefont {Dvornik}}, \ and\ \bibinfo {author}
  {\bibfnamefont {J.}~\bibnamefont {{\r A}kerman}},\ }\href
  {10.1126/sciadv.aax8467} {\bibfield  {journal} {\bibinfo  {journal} {Science
  Advances}\ }\textbf {\bibinfo {volume} {5}},\ \bibinfo {pages} {eaax8467}
  (\bibinfo {year} {2019})}\BibitemShut {NoStop}%
\bibitem [{\citenamefont {Fulara}\ \emph {et~al.}(2020)\citenamefont {Fulara},
  \citenamefont {Zahedinejad}, \citenamefont {Khymyn}, \citenamefont {Dvornik},
  \citenamefont {Fukami}, \citenamefont {Kanai}, \citenamefont {Ohno},\ and\
  \citenamefont {{\AA}kerman}}]{fulara2020natcomm}%
  \BibitemOpen
  \bibfield  {author} {\bibinfo {author} {\bibfnamefont {H.}~\bibnamefont
  {Fulara}}, \bibinfo {author} {\bibfnamefont {M.}~\bibnamefont {Zahedinejad}},
  \bibinfo {author} {\bibfnamefont {R.}~\bibnamefont {Khymyn}}, \bibinfo
  {author} {\bibfnamefont {M.}~\bibnamefont {Dvornik}}, \bibinfo {author}
  {\bibfnamefont {S.}~\bibnamefont {Fukami}}, \bibinfo {author} {\bibfnamefont
  {S.}~\bibnamefont {Kanai}}, \bibinfo {author} {\bibfnamefont
  {H.}~\bibnamefont {Ohno}}, \ and\ \bibinfo {author} {\bibfnamefont
  {J.}~\bibnamefont {{\AA}kerman}},\ }\href
  {https://doi.org/10.1038/s41467-020-17833-x} {\bibfield  {journal} {\bibinfo
  {journal} {Nature Communications}\ }\textbf {\bibinfo {volume} {11}},\
  \bibinfo {pages} {4006} (\bibinfo {year} {2020})}\BibitemShut {NoStop}%
\bibitem [{\citenamefont {Mazraati}\ \emph {et~al.}(2016)\citenamefont
  {Mazraati}, \citenamefont {Chung}, \citenamefont {Houshang}, \citenamefont
  {Dvornik}, \citenamefont {Piazza}, \citenamefont {Qejvanaj}, \citenamefont
  {Jiang}, \citenamefont {Le}, \citenamefont {Weissenrieder},\ and\
  \citenamefont {{\AA}kerman}}]{Mazraati2016apl}%
  \BibitemOpen
  \bibfield  {author} {\bibinfo {author} {\bibfnamefont {H.}~\bibnamefont
  {Mazraati}}, \bibinfo {author} {\bibfnamefont {S.}~\bibnamefont {Chung}},
  \bibinfo {author} {\bibfnamefont {A.}~\bibnamefont {Houshang}}, \bibinfo
  {author} {\bibfnamefont {M.}~\bibnamefont {Dvornik}}, \bibinfo {author}
  {\bibfnamefont {L.}~\bibnamefont {Piazza}}, \bibinfo {author} {\bibfnamefont
  {F.}~\bibnamefont {Qejvanaj}}, \bibinfo {author} {\bibfnamefont
  {S.}~\bibnamefont {Jiang}}, \bibinfo {author} {\bibfnamefont {T.~Q.}\
  \bibnamefont {Le}}, \bibinfo {author} {\bibfnamefont {J.}~\bibnamefont
  {Weissenrieder}}, \ and\ \bibinfo {author} {\bibfnamefont {J.}~\bibnamefont
  {{\AA}kerman}},\ }\href {\doibase 10.1063/1.4971828} {\bibfield  {journal}
  {\bibinfo  {journal} {Appl. Phys. Lett.}\ }\textbf {\bibinfo {volume}
  {109}},\ \bibinfo {pages} {242402} (\bibinfo {year} {2016})}\BibitemShut
  {NoStop}%
\bibitem [{\citenamefont {Evelt}\ \emph {et~al.}(2018)\citenamefont {Evelt},
  \citenamefont {Safranski}, \citenamefont {Aldosary}, \citenamefont {Demidov},
  \citenamefont {Barsukov}, \citenamefont {Nosov}, \citenamefont {Rinkevich},
  \citenamefont {Sobotkiewich}, \citenamefont {Li}, \citenamefont {Shi},
  \citenamefont {Krivorotov},\ and\ \citenamefont
  {Demokritov}}]{Evelt2018scirep}%
  \BibitemOpen
  \bibfield  {author} {\bibinfo {author} {\bibfnamefont {M.}~\bibnamefont
  {Evelt}}, \bibinfo {author} {\bibfnamefont {C.}~\bibnamefont {Safranski}},
  \bibinfo {author} {\bibfnamefont {M.}~\bibnamefont {Aldosary}}, \bibinfo
  {author} {\bibfnamefont {V.~E.}\ \bibnamefont {Demidov}}, \bibinfo {author}
  {\bibfnamefont {I.}~\bibnamefont {Barsukov}}, \bibinfo {author}
  {\bibfnamefont {A.~P.}\ \bibnamefont {Nosov}}, \bibinfo {author}
  {\bibfnamefont {A.~B.}\ \bibnamefont {Rinkevich}}, \bibinfo {author}
  {\bibfnamefont {K.}~\bibnamefont {Sobotkiewich}}, \bibinfo {author}
  {\bibfnamefont {X.}~\bibnamefont {Li}}, \bibinfo {author} {\bibfnamefont
  {J.}~\bibnamefont {Shi}}, \bibinfo {author} {\bibfnamefont {I.~N.}\
  \bibnamefont {Krivorotov}}, \ and\ \bibinfo {author} {\bibfnamefont {S.~O.}\
  \bibnamefont {Demokritov}},\ }\href {\doibase 10.1038/s41598-018-19606-5}
  {\bibfield  {journal} {\bibinfo  {journal} {Scientific Reports}\ }\textbf
  {\bibinfo {volume} {8}},\ \bibinfo {pages} {1269} (\bibinfo {year}
  {2018})}\BibitemShut {NoStop}%
\bibitem [{\citenamefont {Hache}\ \emph {et~al.}(2020)\citenamefont {Hache},
  \citenamefont {Li}, \citenamefont {Weinhold}, \citenamefont {Scheumann},
  \citenamefont {Gonçalves}, \citenamefont {Hellwig}, \citenamefont
  {Fassbender},\ and\ \citenamefont {Schultheiss}}]{Hache2020apl}%
  \BibitemOpen
  \bibfield  {author} {\bibinfo {author} {\bibfnamefont {T.}~\bibnamefont
  {Hache}}, \bibinfo {author} {\bibfnamefont {Y.}~\bibnamefont {Li}}, \bibinfo
  {author} {\bibfnamefont {T.}~\bibnamefont {Weinhold}}, \bibinfo {author}
  {\bibfnamefont {B.}~\bibnamefont {Scheumann}}, \bibinfo {author}
  {\bibfnamefont {F.~J.~T.}\ \bibnamefont {Gonçalves}}, \bibinfo {author}
  {\bibfnamefont {O.}~\bibnamefont {Hellwig}}, \bibinfo {author} {\bibfnamefont
  {J.}~\bibnamefont {Fassbender}}, \ and\ \bibinfo {author} {\bibfnamefont
  {H.}~\bibnamefont {Schultheiss}},\ }\href {\doibase 10.1063/5.0008988}
  {\bibfield  {journal} {\bibinfo  {journal} {Applied Physics Letters}\
  }\textbf {\bibinfo {volume} {116}},\ \bibinfo {pages} {192405} (\bibinfo
  {year} {2020})},\ \Eprint
  {http://arxiv.org/abs/https://doi.org/10.1063/5.0008988}
  {https://doi.org/10.1063/5.0008988} \BibitemShut {NoStop}%
\bibitem [{\citenamefont {Haidar}\ \emph {et~al.}(2021)\citenamefont {Haidar},
  \citenamefont {Mazraati}, \citenamefont {Dürrenfeld}, \citenamefont
  {Fulara}, \citenamefont {Ranjbar},\ and\ \citenamefont
  {Åkerman}}]{Haidar2021apl}%
  \BibitemOpen
  \bibfield  {author} {\bibinfo {author} {\bibfnamefont {M.}~\bibnamefont
  {Haidar}}, \bibinfo {author} {\bibfnamefont {H.}~\bibnamefont {Mazraati}},
  \bibinfo {author} {\bibfnamefont {P.}~\bibnamefont {Dürrenfeld}}, \bibinfo
  {author} {\bibfnamefont {H.}~\bibnamefont {Fulara}}, \bibinfo {author}
  {\bibfnamefont {M.}~\bibnamefont {Ranjbar}}, \ and\ \bibinfo {author}
  {\bibfnamefont {J.}~\bibnamefont {Åkerman}},\ }\href {\doibase
  10.1063/5.0035697} {\bibfield  {journal} {\bibinfo  {journal} {Applied
  Physics Letters}\ }\textbf {\bibinfo {volume} {118}},\ \bibinfo {pages}
  {012406} (\bibinfo {year} {2021})},\ \Eprint
  {http://arxiv.org/abs/https://doi.org/10.1063/5.0035697}
  {https://doi.org/10.1063/5.0035697} \BibitemShut {NoStop}%
\bibitem [{\citenamefont {Sato}\ \emph {et~al.}(2019)\citenamefont {Sato},
  \citenamefont {Schultheiss}, \citenamefont {K\"orber}, \citenamefont
  {Puwenberg}, \citenamefont {M\"uhl}, \citenamefont {Awad}, \citenamefont
  {Arekapudi}, \citenamefont {Hellwig}, \citenamefont {Fassbender},\ and\
  \citenamefont {Schultheiss}}]{Sato2019prl}%
  \BibitemOpen
  \bibfield  {author} {\bibinfo {author} {\bibfnamefont {N.}~\bibnamefont
  {Sato}}, \bibinfo {author} {\bibfnamefont {K.}~\bibnamefont {Schultheiss}},
  \bibinfo {author} {\bibfnamefont {L.}~\bibnamefont {K\"orber}}, \bibinfo
  {author} {\bibfnamefont {N.}~\bibnamefont {Puwenberg}}, \bibinfo {author}
  {\bibfnamefont {T.}~\bibnamefont {M\"uhl}}, \bibinfo {author} {\bibfnamefont
  {A.~A.}\ \bibnamefont {Awad}}, \bibinfo {author} {\bibfnamefont {S.~S.
  P.~K.}\ \bibnamefont {Arekapudi}}, \bibinfo {author} {\bibfnamefont
  {O.}~\bibnamefont {Hellwig}}, \bibinfo {author} {\bibfnamefont
  {J.}~\bibnamefont {Fassbender}}, \ and\ \bibinfo {author} {\bibfnamefont
  {H.}~\bibnamefont {Schultheiss}},\ }\href {\doibase
  10.1103/PhysRevLett.123.057204} {\bibfield  {journal} {\bibinfo  {journal}
  {Phys. Rev. Lett.}\ }\textbf {\bibinfo {volume} {123}},\ \bibinfo {pages}
  {057204} (\bibinfo {year} {2019})}\BibitemShut {NoStop}%
\bibitem [{\citenamefont {Safranski}\ \emph {et~al.}(2019)\citenamefont
  {Safranski}, \citenamefont {Montoya},\ and\ \citenamefont
  {Krivorotov}}]{Safranski2019natnano}%
  \BibitemOpen
  \bibfield  {author} {\bibinfo {author} {\bibfnamefont {C.}~\bibnamefont
  {Safranski}}, \bibinfo {author} {\bibfnamefont {E.~A.}\ \bibnamefont
  {Montoya}}, \ and\ \bibinfo {author} {\bibfnamefont {I.~N.}\ \bibnamefont
  {Krivorotov}},\ }\href {\doibase 10.1038/s41565-018-0282-0} {\bibfield
  {journal} {\bibinfo  {journal} {Nature Nanotechnology}\ }\textbf {\bibinfo
  {volume} {14}},\ \bibinfo {pages} {27} (\bibinfo {year} {2019})}\BibitemShut
  {NoStop}%
\bibitem [{\citenamefont {Chen}\ \emph {et~al.}(2020)\citenamefont {Chen},
  \citenamefont {Smith}, \citenamefont {Montoya}, \citenamefont {Lu},\ and\
  \citenamefont {Krivorotov}}]{Chen2020commphys}%
  \BibitemOpen
  \bibfield  {author} {\bibinfo {author} {\bibfnamefont {J.-R.}\ \bibnamefont
  {Chen}}, \bibinfo {author} {\bibfnamefont {A.}~\bibnamefont {Smith}},
  \bibinfo {author} {\bibfnamefont {E.~A.}\ \bibnamefont {Montoya}}, \bibinfo
  {author} {\bibfnamefont {J.~G.}\ \bibnamefont {Lu}}, \ and\ \bibinfo {author}
  {\bibfnamefont {I.~N.}\ \bibnamefont {Krivorotov}},\ }\href {\doibase
  10.1038/s42005-020-00454-7} {\bibfield  {journal} {\bibinfo  {journal}
  {Communications Physics}\ }\textbf {\bibinfo {volume} {3}},\ \bibinfo {pages}
  {187} (\bibinfo {year} {2020})}\BibitemShut {NoStop}%
\bibitem [{\citenamefont {Awad}\ \emph {et~al.}(2017)\citenamefont {Awad},
  \citenamefont {D{\"{u}}rrenfeld}, \citenamefont {Houshang}, \citenamefont
  {Dvornik}, \citenamefont {Iacocca}, \citenamefont {Dumas},\ and\
  \citenamefont {{\AA}kerman}}]{Awad2016natphys}%
  \BibitemOpen
  \bibfield  {author} {\bibinfo {author} {\bibfnamefont {A.~A.}\ \bibnamefont
  {Awad}}, \bibinfo {author} {\bibfnamefont {P.}~\bibnamefont
  {D{\"{u}}rrenfeld}}, \bibinfo {author} {\bibfnamefont {A.}~\bibnamefont
  {Houshang}}, \bibinfo {author} {\bibfnamefont {M.}~\bibnamefont {Dvornik}},
  \bibinfo {author} {\bibfnamefont {E.}~\bibnamefont {Iacocca}}, \bibinfo
  {author} {\bibfnamefont {R.~K.}\ \bibnamefont {Dumas}}, \ and\ \bibinfo
  {author} {\bibfnamefont {J.}~\bibnamefont {{\AA}kerman}},\ }\href {\doibase
  10.1038/nphys3927} {\bibfield  {journal} {\bibinfo  {journal} {Nat. Phys.}\
  }\textbf {\bibinfo {volume} {13}},\ \bibinfo {pages} {292} (\bibinfo {year}
  {2017})}\BibitemShut {NoStop}%
\bibitem [{\citenamefont {Zahedinejad}\ \emph
  {et~al.}(2020{\natexlab{a}})\citenamefont {Zahedinejad}, \citenamefont
  {Awad}, \citenamefont {Muralidhar}, \citenamefont {Khymyn}, \citenamefont
  {Fulara}, \citenamefont {Mazraati}, \citenamefont {Dvornik},\ and\
  \citenamefont {{\AA}kerman}}]{Zahedinejad2020natnano}%
  \BibitemOpen
  \bibfield  {author} {\bibinfo {author} {\bibfnamefont {M.}~\bibnamefont
  {Zahedinejad}}, \bibinfo {author} {\bibfnamefont {A.~A.}\ \bibnamefont
  {Awad}}, \bibinfo {author} {\bibfnamefont {S.}~\bibnamefont {Muralidhar}},
  \bibinfo {author} {\bibfnamefont {R.}~\bibnamefont {Khymyn}}, \bibinfo
  {author} {\bibfnamefont {H.}~\bibnamefont {Fulara}}, \bibinfo {author}
  {\bibfnamefont {H.}~\bibnamefont {Mazraati}}, \bibinfo {author}
  {\bibfnamefont {M.}~\bibnamefont {Dvornik}}, \ and\ \bibinfo {author}
  {\bibfnamefont {J.}~\bibnamefont {{\AA}kerman}},\ }\href {\doibase
  10.1038/s41565-019-0593-9} {\bibfield  {journal} {\bibinfo  {journal} {Nature
  Nanotechnology}\ }\textbf {\bibinfo {volume} {15}},\ \bibinfo {pages} {47}
  (\bibinfo {year} {2020}{\natexlab{a}})}\BibitemShut {NoStop}%
\bibitem [{\citenamefont {Singh}\ \emph {et~al.}(2021)\citenamefont {Singh},
  \citenamefont {Garg}, \citenamefont {Kumar}, \citenamefont {Muduli},\ and\
  \citenamefont {Bhowmik}}]{Singh2021aipadv}%
  \BibitemOpen
  \bibfield  {author} {\bibinfo {author} {\bibfnamefont {U.}~\bibnamefont
  {Singh}}, \bibinfo {author} {\bibfnamefont {N.}~\bibnamefont {Garg}},
  \bibinfo {author} {\bibfnamefont {S.}~\bibnamefont {Kumar}}, \bibinfo
  {author} {\bibfnamefont {P.~K.}\ \bibnamefont {Muduli}}, \ and\ \bibinfo
  {author} {\bibfnamefont {D.}~\bibnamefont {Bhowmik}},\ }\href {\doibase
  10.1063/9.0000192} {\bibfield  {journal} {\bibinfo  {journal} {AIP Advances}\
  }\textbf {\bibinfo {volume} {11}},\ \bibinfo {pages} {045117} (\bibinfo
  {year} {2021})},\ \Eprint
  {http://arxiv.org/abs/https://doi.org/10.1063/9.0000192}
  {https://doi.org/10.1063/9.0000192} \BibitemShut {NoStop}%
\bibitem [{\citenamefont {Zahedinejad}\ \emph
  {et~al.}(2020{\natexlab{b}})\citenamefont {Zahedinejad}, \citenamefont
  {Fulara}, \citenamefont {Khymyn}, \citenamefont {Houshang}, \citenamefont
  {Fukami}, \citenamefont {Kanai}, \citenamefont {Ohno},\ and\ \citenamefont
  {Åkerman}}]{zahedinejad2020arxiv}%
  \BibitemOpen
  \bibfield  {author} {\bibinfo {author} {\bibfnamefont {M.}~\bibnamefont
  {Zahedinejad}}, \bibinfo {author} {\bibfnamefont {H.}~\bibnamefont {Fulara}},
  \bibinfo {author} {\bibfnamefont {R.}~\bibnamefont {Khymyn}}, \bibinfo
  {author} {\bibfnamefont {A.}~\bibnamefont {Houshang}}, \bibinfo {author}
  {\bibfnamefont {S.}~\bibnamefont {Fukami}}, \bibinfo {author} {\bibfnamefont
  {S.}~\bibnamefont {Kanai}}, \bibinfo {author} {\bibfnamefont
  {H.}~\bibnamefont {Ohno}}, \ and\ \bibinfo {author} {\bibfnamefont
  {J.}~\bibnamefont {Åkerman}},\ }\href@noop {} {\bibfield  {journal}
  {\bibinfo  {journal} {arXiv}\ } (\bibinfo {year} {2020}{\natexlab{b}})},\
  \Eprint {http://arxiv.org/abs/2009.06594} {arXiv:2009.06594 [physics.app-ph]}
  \BibitemShut {NoStop}%
\bibitem [{\citenamefont {Garg}\ \emph {et~al.}(2021)\citenamefont {Garg},
  \citenamefont {Hemadri~Bhotla}, \citenamefont {Muduli},\ and\ \citenamefont
  {Bhowmik}}]{Garg2021neurcomp}%
  \BibitemOpen
  \bibfield  {author} {\bibinfo {author} {\bibfnamefont {N.}~\bibnamefont
  {Garg}}, \bibinfo {author} {\bibfnamefont {S.~V.}\ \bibnamefont
  {Hemadri~Bhotla}}, \bibinfo {author} {\bibfnamefont {P.~K.}\ \bibnamefont
  {Muduli}}, \ and\ \bibinfo {author} {\bibfnamefont {D.}~\bibnamefont
  {Bhowmik}},\ }\href
  {http://iopscience.iop.org/article/10.1088/2634-4386/ac3258} {\bibfield
  {journal} {\bibinfo  {journal} {Neuromorphic Computing and Engineering}\ }
  (\bibinfo {year} {2021})}\BibitemShut {NoStop}%
\bibitem [{\citenamefont {Albertsson}\ \emph {et~al.}(2021)\citenamefont
  {Albertsson}, \citenamefont {Zahedinejad}, \citenamefont {Houshang},
  \citenamefont {Khymyn}, \citenamefont {Åkerman},\ and\ \citenamefont
  {Rusu}}]{Albertsson2021apl}%
  \BibitemOpen
  \bibfield  {author} {\bibinfo {author} {\bibfnamefont {D.~I.}\ \bibnamefont
  {Albertsson}}, \bibinfo {author} {\bibfnamefont {M.}~\bibnamefont
  {Zahedinejad}}, \bibinfo {author} {\bibfnamefont {A.}~\bibnamefont
  {Houshang}}, \bibinfo {author} {\bibfnamefont {R.}~\bibnamefont {Khymyn}},
  \bibinfo {author} {\bibfnamefont {J.}~\bibnamefont {Åkerman}}, \ and\
  \bibinfo {author} {\bibfnamefont {A.}~\bibnamefont {Rusu}},\ }\href {\doibase
  10.1063/5.0041575} {\bibfield  {journal} {\bibinfo  {journal} {Applied
  Physics Letters}\ }\textbf {\bibinfo {volume} {118}},\ \bibinfo {pages}
  {112404} (\bibinfo {year} {2021})},\ \Eprint
  {http://arxiv.org/abs/https://doi.org/10.1063/5.0041575}
  {https://doi.org/10.1063/5.0041575} \BibitemShut {NoStop}%
\bibitem [{\citenamefont {Houshang}\ \emph {et~al.}(2020)\citenamefont
  {Houshang}, \citenamefont {Zahedinejad}, \citenamefont {Muralidhar},
  \citenamefont {Checinski}, \citenamefont {Awad},\ and\ \citenamefont
  {Åkerman}}]{houshang2020arxiv}%
  \BibitemOpen
  \bibfield  {author} {\bibinfo {author} {\bibfnamefont {A.}~\bibnamefont
  {Houshang}}, \bibinfo {author} {\bibfnamefont {M.}~\bibnamefont
  {Zahedinejad}}, \bibinfo {author} {\bibfnamefont {S.}~\bibnamefont
  {Muralidhar}}, \bibinfo {author} {\bibfnamefont {J.}~\bibnamefont
  {Checinski}}, \bibinfo {author} {\bibfnamefont {A.~A.}\ \bibnamefont {Awad}},
  \ and\ \bibinfo {author} {\bibfnamefont {J.}~\bibnamefont {Åkerman}},\
  }\href@noop {} {\bibfield  {journal} {\bibinfo  {journal} {arXiv}\ }
  (\bibinfo {year} {2020})},\ \Eprint {http://arxiv.org/abs/2006.02236}
  {arXiv:2006.02236 [cond-mat.mes-hall]} \BibitemShut {NoStop}%
\bibitem [{\citenamefont {Pai}\ \emph {et~al.}(2012)\citenamefont {Pai},
  \citenamefont {Liu}, \citenamefont {Li}, \citenamefont {Tseng}, \citenamefont
  {Ralph},\ and\ \citenamefont {Buhrman}}]{Pai2012}%
  \BibitemOpen
  \bibfield  {author} {\bibinfo {author} {\bibfnamefont {C.-F.}\ \bibnamefont
  {Pai}}, \bibinfo {author} {\bibfnamefont {L.}~\bibnamefont {Liu}}, \bibinfo
  {author} {\bibfnamefont {Y.}~\bibnamefont {Li}}, \bibinfo {author}
  {\bibfnamefont {H.}~\bibnamefont {Tseng}}, \bibinfo {author} {\bibfnamefont
  {D.}~\bibnamefont {Ralph}}, \ and\ \bibinfo {author} {\bibfnamefont {R.~A.}\
  \bibnamefont {Buhrman}},\ }\href {\doibase 10.1063/1.4753947} {\bibfield
  {journal} {\bibinfo  {journal} {Applied Physics Letters}\ }\textbf {\bibinfo
  {volume} {101}},\ \bibinfo {pages} {122404} (\bibinfo {year}
  {2012})}\BibitemShut {NoStop}%
\bibitem [{\citenamefont {Demasius}\ \emph {et~al.}(2016)\citenamefont
  {Demasius}, \citenamefont {Phung}, \citenamefont {Zhang}, \citenamefont
  {Hughes}, \citenamefont {Yang}, \citenamefont {Kellock}, \citenamefont {Han},
  \citenamefont {Pushp},\ and\ \citenamefont {Parkin}}]{Demasius2016natcomm}%
  \BibitemOpen
  \bibfield  {author} {\bibinfo {author} {\bibfnamefont {K.~U.}\ \bibnamefont
  {Demasius}}, \bibinfo {author} {\bibfnamefont {T.}~\bibnamefont {Phung}},
  \bibinfo {author} {\bibfnamefont {W.}~\bibnamefont {Zhang}}, \bibinfo
  {author} {\bibfnamefont {B.~P.}\ \bibnamefont {Hughes}}, \bibinfo {author}
  {\bibfnamefont {S.~H.}\ \bibnamefont {Yang}}, \bibinfo {author}
  {\bibfnamefont {A.}~\bibnamefont {Kellock}}, \bibinfo {author} {\bibfnamefont
  {W.}~\bibnamefont {Han}}, \bibinfo {author} {\bibfnamefont {A.}~\bibnamefont
  {Pushp}}, \ and\ \bibinfo {author} {\bibfnamefont {S.~S.~P.}\ \bibnamefont
  {Parkin}},\ }\href {\doibase https://doi.org/10.1038/ncomms10644} {\bibfield
  {journal} {\bibinfo  {journal} {Nat. Commun.}\ }\textbf {\bibinfo {volume}
  {7}},\ \bibinfo {pages} {10644} (\bibinfo {year} {2016})}\BibitemShut
  {NoStop}%
\bibitem [{\citenamefont {Shashank}\ \emph {et~al.}(2021)\citenamefont
  {Shashank}, \citenamefont {Medwal}, \citenamefont {Nakamura}, \citenamefont
  {Mohan}, \citenamefont {Nongjai}, \citenamefont {Kandasami}, \citenamefont
  {Rawat}, \citenamefont {Asada}, \citenamefont {Gupta},\ and\ \citenamefont
  {Fukuma}}]{utkarsh2021apl}%
  \BibitemOpen
  \bibfield  {author} {\bibinfo {author} {\bibfnamefont {U.}~\bibnamefont
  {Shashank}}, \bibinfo {author} {\bibfnamefont {R.}~\bibnamefont {Medwal}},
  \bibinfo {author} {\bibfnamefont {Y.}~\bibnamefont {Nakamura}}, \bibinfo
  {author} {\bibfnamefont {J.~R.}\ \bibnamefont {Mohan}}, \bibinfo {author}
  {\bibfnamefont {R.}~\bibnamefont {Nongjai}}, \bibinfo {author} {\bibfnamefont
  {A.}~\bibnamefont {Kandasami}}, \bibinfo {author} {\bibfnamefont {R.~S.}\
  \bibnamefont {Rawat}}, \bibinfo {author} {\bibfnamefont {H.}~\bibnamefont
  {Asada}}, \bibinfo {author} {\bibfnamefont {S.}~\bibnamefont {Gupta}}, \ and\
  \bibinfo {author} {\bibfnamefont {Y.}~\bibnamefont {Fukuma}},\ }\href
  {\doibase 10.1063/5.0054779} {\bibfield  {journal} {\bibinfo  {journal}
  {Applied Physics Letters}\ }\textbf {\bibinfo {volume} {118}},\ \bibinfo
  {pages} {252406} (\bibinfo {year} {2021})},\ \Eprint
  {http://arxiv.org/abs/https://doi.org/10.1063/5.0054779}
  {https://doi.org/10.1063/5.0054779} \BibitemShut {NoStop}%
\bibitem [{\citenamefont {Nguyen}\ \emph {et~al.}(2015)\citenamefont {Nguyen},
  \citenamefont {Pai}, \citenamefont {Nguyen}, \citenamefont {Muller},
  \citenamefont {Ralph},\ and\ \citenamefont {Buhrman}}]{nguyen2015apl}%
  \BibitemOpen
  \bibfield  {author} {\bibinfo {author} {\bibfnamefont {M.-H.}\ \bibnamefont
  {Nguyen}}, \bibinfo {author} {\bibfnamefont {C.-F.}\ \bibnamefont {Pai}},
  \bibinfo {author} {\bibfnamefont {K.~X.}\ \bibnamefont {Nguyen}}, \bibinfo
  {author} {\bibfnamefont {D.~A.}\ \bibnamefont {Muller}}, \bibinfo {author}
  {\bibfnamefont {D.~C.}\ \bibnamefont {Ralph}}, \ and\ \bibinfo {author}
  {\bibfnamefont {R.~A.}\ \bibnamefont {Buhrman}},\ }\href {\doibase
  10.1063/1.4922084} {\bibfield  {journal} {\bibinfo  {journal} {Applied
  Physics Letters}\ }\textbf {\bibinfo {volume} {106}},\ \bibinfo {pages}
  {222402} (\bibinfo {year} {2015})},\ \Eprint
  {http://arxiv.org/abs/https://doi.org/10.1063/1.4922084}
  {https://doi.org/10.1063/1.4922084} \BibitemShut {NoStop}%
\bibitem [{\citenamefont {Mazraati}\ \emph
  {et~al.}(2018{\natexlab{b}})\citenamefont {Mazraati}, \citenamefont
  {Zahedinejad},\ and\ \citenamefont {Åkerman}}]{mazraati2018apl}%
  \BibitemOpen
  \bibfield  {author} {\bibinfo {author} {\bibfnamefont {H.}~\bibnamefont
  {Mazraati}}, \bibinfo {author} {\bibfnamefont {M.}~\bibnamefont
  {Zahedinejad}}, \ and\ \bibinfo {author} {\bibfnamefont {J.}~\bibnamefont
  {Åkerman}},\ }\href {\doibase 10.1063/1.5026232} {\bibfield  {journal}
  {\bibinfo  {journal} {Applied Physics Letters}\ }\textbf {\bibinfo {volume}
  {113}},\ \bibinfo {pages} {092401} (\bibinfo {year}
  {2018}{\natexlab{b}})}\BibitemShut {NoStop}%
\bibitem [{\citenamefont {Obstbaum}\ \emph {et~al.}(2016)\citenamefont
  {Obstbaum}, \citenamefont {Decker}, \citenamefont {Greitner}, \citenamefont
  {Haertinger}, \citenamefont {Meier}, \citenamefont {Kronseder}, \citenamefont
  {Chadova}, \citenamefont {Wimmer}, \citenamefont {K\"odderitzsch},
  \citenamefont {Ebert},\ and\ \citenamefont {Back}}]{obstbaum2016prl}%
  \BibitemOpen
  \bibfield  {author} {\bibinfo {author} {\bibfnamefont {M.}~\bibnamefont
  {Obstbaum}}, \bibinfo {author} {\bibfnamefont {M.}~\bibnamefont {Decker}},
  \bibinfo {author} {\bibfnamefont {A.~K.}\ \bibnamefont {Greitner}}, \bibinfo
  {author} {\bibfnamefont {M.}~\bibnamefont {Haertinger}}, \bibinfo {author}
  {\bibfnamefont {T.~N.~G.}\ \bibnamefont {Meier}}, \bibinfo {author}
  {\bibfnamefont {M.}~\bibnamefont {Kronseder}}, \bibinfo {author}
  {\bibfnamefont {K.}~\bibnamefont {Chadova}}, \bibinfo {author} {\bibfnamefont
  {S.}~\bibnamefont {Wimmer}}, \bibinfo {author} {\bibfnamefont
  {D.}~\bibnamefont {K\"odderitzsch}}, \bibinfo {author} {\bibfnamefont
  {H.}~\bibnamefont {Ebert}}, \ and\ \bibinfo {author} {\bibfnamefont {C.~H.}\
  \bibnamefont {Back}},\ }\href {\doibase 10.1103/PhysRevLett.117.167204}
  {\bibfield  {journal} {\bibinfo  {journal} {Phys. Rev. Lett.}\ }\textbf
  {\bibinfo {volume} {117}},\ \bibinfo {pages} {167204} (\bibinfo {year}
  {2016})}\BibitemShut {NoStop}%
\bibitem [{\citenamefont {Sui}\ \emph {et~al.}(2017)\citenamefont {Sui},
  \citenamefont {Wang}, \citenamefont {Kim}, \citenamefont {Wang},
  \citenamefont {Rhim}, \citenamefont {Duan},\ and\ \citenamefont
  {Kioussis}}]{sui2017prb}%
  \BibitemOpen
  \bibfield  {author} {\bibinfo {author} {\bibfnamefont {X.}~\bibnamefont
  {Sui}}, \bibinfo {author} {\bibfnamefont {C.}~\bibnamefont {Wang}}, \bibinfo
  {author} {\bibfnamefont {J.}~\bibnamefont {Kim}}, \bibinfo {author}
  {\bibfnamefont {J.}~\bibnamefont {Wang}}, \bibinfo {author} {\bibfnamefont
  {S.~H.}\ \bibnamefont {Rhim}}, \bibinfo {author} {\bibfnamefont
  {W.}~\bibnamefont {Duan}}, \ and\ \bibinfo {author} {\bibfnamefont
  {N.}~\bibnamefont {Kioussis}},\ }\href {\doibase 10.1103/PhysRevB.96.241105}
  {\bibfield  {journal} {\bibinfo  {journal} {Phys. Rev. B}\ }\textbf {\bibinfo
  {volume} {96}},\ \bibinfo {pages} {241105} (\bibinfo {year}
  {2017})}\BibitemShut {NoStop}%
\bibitem [{\citenamefont {Zhu}\ \emph {et~al.}(2018)\citenamefont {Zhu},
  \citenamefont {Ralph},\ and\ \citenamefont {Buhrman}}]{zhu2018pra}%
  \BibitemOpen
  \bibfield  {author} {\bibinfo {author} {\bibfnamefont {L.}~\bibnamefont
  {Zhu}}, \bibinfo {author} {\bibfnamefont {D.~C.}\ \bibnamefont {Ralph}}, \
  and\ \bibinfo {author} {\bibfnamefont {R.~A.}\ \bibnamefont {Buhrman}},\
  }\href {\doibase 10.1103/PhysRevApplied.10.031001} {\bibfield  {journal}
  {\bibinfo  {journal} {Phys. Rev. Applied}\ }\textbf {\bibinfo {volume}
  {10}},\ \bibinfo {pages} {031001} (\bibinfo {year} {2018})}\BibitemShut
  {NoStop}%
\bibitem [{\citenamefont {Kim}\ \emph {et~al.}(2020)\citenamefont {Kim},
  \citenamefont {Han}, \citenamefont {Vafaee}, \citenamefont {Jaiswal},
  \citenamefont {Lee}, \citenamefont {Jakob},\ and\ \citenamefont
  {Kläui}}]{kim2020apl}%
  \BibitemOpen
  \bibfield  {author} {\bibinfo {author} {\bibfnamefont {J.-Y.}\ \bibnamefont
  {Kim}}, \bibinfo {author} {\bibfnamefont {D.-S.}\ \bibnamefont {Han}},
  \bibinfo {author} {\bibfnamefont {M.}~\bibnamefont {Vafaee}}, \bibinfo
  {author} {\bibfnamefont {S.}~\bibnamefont {Jaiswal}}, \bibinfo {author}
  {\bibfnamefont {K.}~\bibnamefont {Lee}}, \bibinfo {author} {\bibfnamefont
  {G.}~\bibnamefont {Jakob}}, \ and\ \bibinfo {author} {\bibfnamefont
  {M.}~\bibnamefont {Kläui}},\ }\href {\doibase 10.1063/5.0022012} {\bibfield
  {journal} {\bibinfo  {journal} {Applied Physics Letters}\ }\textbf {\bibinfo
  {volume} {117}},\ \bibinfo {pages} {142403} (\bibinfo {year} {2020})},\
  \Eprint {http://arxiv.org/abs/https://doi.org/10.1063/5.0022012}
  {https://doi.org/10.1063/5.0022012} \BibitemShut {NoStop}%
\bibitem [{\citenamefont {Skinner}\ \emph {et~al.}(2014)\citenamefont
  {Skinner}, \citenamefont {Wang}, \citenamefont {Hindmarch}, \citenamefont
  {Rushforth}, \citenamefont {Irvine}, \citenamefont {Heiss}, \citenamefont
  {Kurebayashi},\ and\ \citenamefont {Ferguson}}]{skinner2014apl}%
  \BibitemOpen
  \bibfield  {author} {\bibinfo {author} {\bibfnamefont {T.}~\bibnamefont
  {Skinner}}, \bibinfo {author} {\bibfnamefont {M.}~\bibnamefont {Wang}},
  \bibinfo {author} {\bibfnamefont {A.}~\bibnamefont {Hindmarch}}, \bibinfo
  {author} {\bibfnamefont {A.}~\bibnamefont {Rushforth}}, \bibinfo {author}
  {\bibfnamefont {A.}~\bibnamefont {Irvine}}, \bibinfo {author} {\bibfnamefont
  {D.}~\bibnamefont {Heiss}}, \bibinfo {author} {\bibfnamefont
  {H.}~\bibnamefont {Kurebayashi}}, \ and\ \bibinfo {author} {\bibfnamefont
  {A.}~\bibnamefont {Ferguson}},\ }\href {\doibase 10.1063/1.4864399}
  {\bibfield  {journal} {\bibinfo  {journal} {Applied Physics Letters}\
  }\textbf {\bibinfo {volume} {104}},\ \bibinfo {pages} {062401} (\bibinfo
  {year} {2014})}\BibitemShut {NoStop}%
\bibitem [{\citenamefont {Petroff}\ \emph {et~al.}(1973)\citenamefont
  {Petroff}, \citenamefont {Sheng}, \citenamefont {Sinha}, \citenamefont
  {Rozgonyi},\ and\ \citenamefont {Alexander}}]{petroff1973microstructure}%
  \BibitemOpen
  \bibfield  {author} {\bibinfo {author} {\bibfnamefont {P.}~\bibnamefont
  {Petroff}}, \bibinfo {author} {\bibfnamefont {T.}~\bibnamefont {Sheng}},
  \bibinfo {author} {\bibfnamefont {A.}~\bibnamefont {Sinha}}, \bibinfo
  {author} {\bibfnamefont {G.}~\bibnamefont {Rozgonyi}}, \ and\ \bibinfo
  {author} {\bibfnamefont {F.}~\bibnamefont {Alexander}},\ }\href {\doibase
  10.1063/1.1662611} {\bibfield  {journal} {\bibinfo  {journal} {Journal of
  Applied Physics}\ }\textbf {\bibinfo {volume} {44}},\ \bibinfo {pages} {2545}
  (\bibinfo {year} {1973})}\BibitemShut {NoStop}%
\bibitem [{\citenamefont {Liu}\ \emph {et~al.}(2011)\citenamefont {Liu},
  \citenamefont {Moriyama}, \citenamefont {Ralph},\ and\ \citenamefont
  {Buhrman}}]{Liu2011prl}%
  \BibitemOpen
  \bibfield  {author} {\bibinfo {author} {\bibfnamefont {L.}~\bibnamefont
  {Liu}}, \bibinfo {author} {\bibfnamefont {T.}~\bibnamefont {Moriyama}},
  \bibinfo {author} {\bibfnamefont {D.~C.}\ \bibnamefont {Ralph}}, \ and\
  \bibinfo {author} {\bibfnamefont {R.~A.}\ \bibnamefont {Buhrman}},\ }\href
  {\doibase 10.1103/PhysRevLett.106.036601} {\bibfield  {journal} {\bibinfo
  {journal} {Phys. Rev. Lett.}\ }\textbf {\bibinfo {volume} {106}},\ \bibinfo
  {pages} {036601} (\bibinfo {year} {2011})}\BibitemShut {NoStop}%
\bibitem [{\citenamefont {Kim}\ \emph {et~al.}(2013)\citenamefont {Kim},
  \citenamefont {Sinha}, \citenamefont {Hayashi}, \citenamefont {Yamanouchi},
  \citenamefont {Fukami}, \citenamefont {Suzuki}, \citenamefont {Mitani},\ and\
  \citenamefont {Ohno}}]{kim2013natmat}%
  \BibitemOpen
  \bibfield  {author} {\bibinfo {author} {\bibfnamefont {J.}~\bibnamefont
  {Kim}}, \bibinfo {author} {\bibfnamefont {J.}~\bibnamefont {Sinha}}, \bibinfo
  {author} {\bibfnamefont {M.}~\bibnamefont {Hayashi}}, \bibinfo {author}
  {\bibfnamefont {M.}~\bibnamefont {Yamanouchi}}, \bibinfo {author}
  {\bibfnamefont {S.}~\bibnamefont {Fukami}}, \bibinfo {author} {\bibfnamefont
  {T.}~\bibnamefont {Suzuki}}, \bibinfo {author} {\bibfnamefont
  {S.}~\bibnamefont {Mitani}}, \ and\ \bibinfo {author} {\bibfnamefont
  {H.}~\bibnamefont {Ohno}},\ }\href {\doibase 10.1038/nmat3522} {\bibfield
  {journal} {\bibinfo  {journal} {Nature materials}\ }\textbf {\bibinfo
  {volume} {12}},\ \bibinfo {pages} {240} (\bibinfo {year} {2013})}\BibitemShut
  {NoStop}%
\bibitem [{\citenamefont {Takeuchi}\ \emph {et~al.}(2018)\citenamefont
  {Takeuchi}, \citenamefont {Zhang}, \citenamefont {Okada}, \citenamefont
  {Sato}, \citenamefont {Fukami},\ and\ \citenamefont
  {Ohno}}]{takeuchi2018apl}%
  \BibitemOpen
  \bibfield  {author} {\bibinfo {author} {\bibfnamefont {Y.}~\bibnamefont
  {Takeuchi}}, \bibinfo {author} {\bibfnamefont {C.}~\bibnamefont {Zhang}},
  \bibinfo {author} {\bibfnamefont {A.}~\bibnamefont {Okada}}, \bibinfo
  {author} {\bibfnamefont {H.}~\bibnamefont {Sato}}, \bibinfo {author}
  {\bibfnamefont {S.}~\bibnamefont {Fukami}}, \ and\ \bibinfo {author}
  {\bibfnamefont {H.}~\bibnamefont {Ohno}},\ }\href {\doibase
  10.1063/1.5027855} {\bibfield  {journal} {\bibinfo  {journal} {Applied
  Physics Letters}\ }\textbf {\bibinfo {volume} {112}},\ \bibinfo {pages}
  {192408} (\bibinfo {year} {2018})}\BibitemShut {NoStop}%
\bibitem [{\citenamefont {Qian}\ \emph {et~al.}(2020)\citenamefont {Qian},
  \citenamefont {Wang}, \citenamefont {Zheng},\ and\ \citenamefont
  {Xiao}}]{qian2020spin}%
  \BibitemOpen
  \bibfield  {author} {\bibinfo {author} {\bibfnamefont {L.}~\bibnamefont
  {Qian}}, \bibinfo {author} {\bibfnamefont {K.}~\bibnamefont {Wang}}, \bibinfo
  {author} {\bibfnamefont {Y.}~\bibnamefont {Zheng}}, \ and\ \bibinfo {author}
  {\bibfnamefont {G.}~\bibnamefont {Xiao}},\ }\href {\doibase
  10.1103/PhysRevB.102.094438} {\bibfield  {journal} {\bibinfo  {journal}
  {Physical Review B}\ }\textbf {\bibinfo {volume} {102}},\ \bibinfo {pages}
  {094438} (\bibinfo {year} {2020})}\BibitemShut {NoStop}%
\bibitem [{\citenamefont {Derunova}\ \emph {et~al.}(2019)\citenamefont
  {Derunova}, \citenamefont {Sun}, \citenamefont {Felser}, \citenamefont
  {Parkin}, \citenamefont {Yan},\ and\ \citenamefont
  {Ali}}]{derunova2019giant}%
  \BibitemOpen
  \bibfield  {author} {\bibinfo {author} {\bibfnamefont {E.}~\bibnamefont
  {Derunova}}, \bibinfo {author} {\bibfnamefont {Y.}~\bibnamefont {Sun}},
  \bibinfo {author} {\bibfnamefont {C.}~\bibnamefont {Felser}}, \bibinfo
  {author} {\bibfnamefont {S.}~\bibnamefont {Parkin}}, \bibinfo {author}
  {\bibfnamefont {B.}~\bibnamefont {Yan}}, \ and\ \bibinfo {author}
  {\bibfnamefont {M.}~\bibnamefont {Ali}},\ }\href {\doibase
  10.1126/sciadv.aav8575} {\bibfield  {journal} {\bibinfo  {journal} {Science
  advances}\ }\textbf {\bibinfo {volume} {5}},\ \bibinfo {pages} {eaav8575}
  (\bibinfo {year} {2019})}\BibitemShut {NoStop}%
\bibitem [{\citenamefont {{Tiberkevich}}\ \emph {et~al.}(2007)\citenamefont
  {{Tiberkevich}}, \citenamefont {{Slavin}},\ and\ \citenamefont
  {{Kim}}}]{tiberkevich2007apl}%
  \BibitemOpen
  \bibfield  {author} {\bibinfo {author} {\bibfnamefont {V.}~\bibnamefont
  {{Tiberkevich}}}, \bibinfo {author} {\bibfnamefont {A.}~\bibnamefont
  {{Slavin}}}, \ and\ \bibinfo {author} {\bibfnamefont {J.-V.}\ \bibnamefont
  {{Kim}}},\ }\href {\doibase 10.1063/1.2812546} {\bibfield  {journal}
  {\bibinfo  {journal} {Appl. Phys. Lett.}\ }\textbf {\bibinfo {volume} {91}},\
  \bibinfo {pages} {192506} (\bibinfo {year} {2007})}\BibitemShut {NoStop}%
\end{thebibliography}
 
%

\end{document}